\documentclass[12pt,preprint]{aastex}

\def\um{$\mu$m }
\def\h2o{H$_2$O}
\def\ch4{CH$_4$}
\def\arcs{\ifmmode {''}\else $''$\fi}

\accepted{2006 December 11}
\begin{document}

\title{THE NIRSPEC BROWN DWARF SPECTROSCOPIC SURVEY II:
HIGH-RESOLUTION J-BAND SPECTRA OF M, L and T DWARFS\footnote{Data
presented herein were obtained at the W.M. Keck Observatory, which
is operated as a scientific partnership among the California
Institute of Technology, the University of California and the
National Aeronautics and Space Administration.  The Observatory
was made possible by the generous financial support of the W.M.
Keck Foundation.}}

\author{{IAN S. MCLEAN\altaffilmark{1}, L. PRATO\altaffilmark{2},
MARK R. MCGOVERN\altaffilmark{1,3}, ADAM J.
BURGASSER\altaffilmark{4}, J. DAVY KIRKPATRICK\altaffilmark{5},
EMILY L. RICE\altaffilmark{1} AND SUNGSOO S. KIM\altaffilmark{6}}}

\altaffiltext{1}{Department of Physics and Astronomy, UCLA, Los
Angeles, CA 90095-1562; mclean@astro.ucla.edu}
\altaffiltext{2}{Lowell Observatory, 1400 West Mars Hill Road, Flagstaff, AZ 86001}
\altaffiltext{3}{Antelope Valley College, 3041 West Avenue K, Lancaster, CA 93536}
\altaffiltext{4}{Massachusetts Institute of Technology, 77 Massachusets
Avenue, Building 37, Cambridge, MA 02139} 
\altaffiltext{5}{Infrared Processing and Analysis Center, MS 100-22,
  California Institute of Technology, Pasadena, CA 91125}
\altaffiltext{6}{Department of Astronomy \& Space Science, Kyung Hee
University, Yongin-shi, Kyunggi-do 449-701, Korea;
sskim@ap.khu.ac.kr}

\begin{abstract}

We present a sequence of high resolution (R$\sim$20,000 or 15 km
s$^{-1}$) infrared spectra of stars and brown dwarfs spanning
spectral types M2.5 to T6. Observations of 16 objects were
obtained using eight echelle orders to cover part of the $J$-band
from 1.165-1.323 $\mu$m with NIRSPEC on the Keck II telescope. By
comparing opacity plots and line lists, over 200 weak features in
the $J$-band are identified with either FeH or H$_{2}$O
transitions. Absorption by FeH attains maximum strength in the
mid-L dwarfs, while H$_{2}$O absorption becomes systematically
stronger towards later spectral types. Narrow resolved features
broaden markedly after the M to L transition. Our high resolution
spectra also reveal that the disappearance of neutral Al lines at
the boundary between M and L dwarfs is
remarkably abrupt, presumably because of the formation of grains.
Neutral Fe lines can be traced to mid-L dwarfs before Fe is
removed by condensation. The neutral potassium (K I) doublets that
dominate the $J$-band have pressure broadened wings that continue
to broaden from $\sim$ 50 km s$^{-1}$ (FWHM) at mid-M to $\sim$
500 km s$^{-1}$ at mid-T. In contrast however, the measured
pseudo-equivalent widths of these same lines reach a maximum in
the mid-L dwarfs. The young L2 dwarf, G196-3B, exhibits narrow
potassium lines without extensive pressure-broadened wings,
indicative of a lower gravity atmosphere. Kelu-1AB, another L2, has
exceptionally broad infrared lines, including FeH and H$_{2}$O
features, confirming its status as a rapid rotator. In contrast to
other late T objects, the peculiar T6 dwarf 2MASS 0937+29
displays a complete absence of potassium even at high resolution,
which may be a metallicity effect or a result of a cooler,
higher-gravity atmosphere.

\end{abstract}


\section{Introduction}

With effective temperatures $\la$ 2200 K, the cool atmospheres of L
and T dwarfs generate complex spectra that are rich in molecular
features, especially at near-infrared (NIR) wavelengths where
ro-vibrational transitions of many molecules dominate.
Fortunately, these cool, very low luminosity objects are also
brightest in the NIR.  Until recently, most infrared spectroscopic
investigations of L and T dwarfs have concentrated on the
identification of strong, broad spectral features, useful for the
establishment of spectral classification, and have employed
resolving powers of $R = \lambda/\Delta\lambda \la 2000$
(Burgasser et al.\ 2002, 2004, 2006; Cushing et al.\ 2003, 2005; Geballe
et al.\ 1996, 2002; Jones et al.\ 1994; Leggett et al.\ 2000,
 2001; McLean et al.\ 2000a, 2001, 2003a; Reid et al.\ 2000; Testi
et al.\ 2001). Observations with significantly higher spectral
resolution are potentially very important because line-blending
from molecular transitions is reduced and weak features are
resolved. Higher resolution spectra are more useful for
constraining models of the complex molecular chemistry of brown
dwarf atmospheres and for characterizing properties such as
gravity and metallicity (Mohanty et al.\ 2004). For example, less
massive brown dwarfs and younger brown dwarfs have smaller surface
gravities which results in less pressure broadening and a
different line shape. Furthermore, spectra with $R\ga$ 20,000
($\la$15 km s$^{-1}$) are required for the measurement of radial
and rotational velocities, and to search for radial velocity
variability associated with brown dwarf spectroscopic binaries.

Obtaining high signal-to-noise observations with an increase in
spectral resolution of a factor of ten is difficult because brown
dwarfs are so faint. Basri et al.\ (2000) successfully resolved
the resonance absorption lines of Cs and Rb in the far-red visible
regime for a sample of M and L dwarfs using the HIRES echelle
spectrograph on the Keck 10-m telescope and derived effective
temperatures through comparison with available model atmospheres.
Reid et al. (2002) also used high-resolution optical echelle
spectroscopy to study 39 dwarfs with spectral types between M6.5
and L0.5. However, because brown dwarf fluxes are significantly
less in visible light, high-resolution observations of fainter L
dwarfs and of the even dimmer T dwarfs are not tenable in the
optical and require infrared observations. With the advent and
development of sensitive large-format IR array detectors, IR
spectroscopy with the requisite spectral resolution is now
possible (McLean et al.\ 1998, 2000b).

In this paper we present the first well-sampled spectral sequence of late M, L
and T dwarfs observed at high resolution (R $\sim$ 20,000) in the
NIR. This work is part of the NIRSPEC Brown Dwarf Spectroscopic
Survey (BDSS) being carried out at the Keck Observatory;
preliminary results were presented in McLean et al.\ (2003b). The
goals of the BDSS are to obtain a significant sample of NIR
spectra of low-mass stars and brown dwarfs of differing ages,
surface gravities, and metallicities at both medium (R $\sim$ 2,000) and
high spectral resolution for spectral classification studies and
comparisons with model atmospheres. McLean et al.\ (2003a),
hereafter M03, describes the lower resolution part of the survey;
spectra from that study are available
online.\footnote{http://www.astro.ucla.edu/$\sim$mclean/BDSSarchive/}
Here, we investigate the $J$-band using ten times higher spectral
resolution than in M03. The $J$-band (defined as 1.15-1.36 $\mu$m
in this paper) is important because this region contains four
strong lines of neutral potassium (K I) that are both temperature
and gravity-sensitive, and which persist throughout the M, L, and
T dwarf sequence.  In \S 2 we describe our observations and data
reduction procedures. \S 3 provides a discussion of the rich,
spectral morphology. In addition to atomic K I, there are lines of
Al I, Fe I, Mn I, Na I and Ti I, and transitions of molecular
species such as CrH, FeH, and H$_{2}$O that can provide a unique
resource for improving model atmospheres at these low
temperatures. We show that the sudden disappearance of the Al I
lines critically defines the M-L boundary at these resolutions.
Concentrating on the strongly pressure-broadened K I lines, we
look for correlations between spectral type and equivalent widths,
velocity widths (FWHM), and residual intensity. The relation
between molecular line strengths and spectral type is also
investigated. The effects of rotation, surface gravity and
metallicity are explored in \S 4. A summary of the overall results
and concluding remarks is given in \S 5.

\section{Observations and Data Reduction}
\subsection{Targets and Instrumentation}
Targets for the initial survey, the BDSS (M03), were selected
primarily from well-known M dwarfs and from L and T dwarfs
identified in the Two Micron All Sky Survey (2MASS; Kirkpatrick et
al. 1997, 1999, 2000, 2001; Burgasser et al. 2000, 2002; Reid et
al. 2000; Wilson et al. 2001), augmented with discoveries from the
Deep Near-Infrared Survey of the Southern Sky (DENIS; Delfosse et al.
1997, 1999), the Sloan Digital Sky Survey (SDSS; Leggett et al. 2000;
Geballe et al. 2002), and other investigations (Becklin \& Zuckerman
1988; Ruiz, Leggett and Allard 1997). To ensure high signal-to-noise spectra for the
high-resolution part of the survey, a subset of 12 of the
brightest objects ($J=7-15$), spanning the spectral type range
from M6 to T6, was selected. Of these, only 2MASS 0140+27 was
not part of the initial survey. The M2.5 star G196-3A was also
observed along with its L2 companion G196-3B, both examples of
objects substantially younger than 1 Gyr (Rebolo et al. 1998) and
therefore most likely to exhibit gravity effects (McGovern et al.
2004). In addition, the peculiar T6 dwarf 2MASS 0937+29
(Burgasser et al. 2002) was added to the list because of its
apparent lack of K I features in lower resolution spectra. Another
late T dwarf (2MASS 2356$-$15, T5.5) was observed after completion
of the initial set for comparison to the 2MASS 0937+29. Although
the signal-to-noise ratio was sufficient to establish the presence
of K I absorption in this T5.5 dwarf, the fainter magnitude
($J=15.8$) and stronger \h2o absorption made quantitative analysis
too difficult so the spectrum is not shown.

Table 1 provides the complete list of 16 targets and the observing
log. Shorthand names such as 2MASS 1507$-$16 are used in the text
for simplicity, but the full designations are listed in Table 1.
Two targets were known visual doubles at the time of observing,
2MASS 0746+20 (L0.5) and DENIS 0205$-$11 (L7), but in neither
case did we have sufficient angular resolution to separate the
components. Subsequent to making our observations, DENIS 0205-11
was reported as a possible triple brown dwarf system (Bouy et al.
2005) based on Hubble Telescope images. Burgasser et al. (2005)
subsequently found SDSS 0423$-$04 to be double, the average
spectral type of T0 being due to an L6 and T2 combination. Even
more recently, the binary nature of Kelu-1, a 0$\farcs$29 pair, was
revealed using Laser Guide Star adaptive optics on the Keck
telescope (Liu \& Leggett 2005; Gelino et al. 2006). Again, in neither
case were these targets resolved in our NIRSPEC observations.

All of the observations were made using the NIRSPEC cryogenic
spectrometer on the Keck II 10-m telescope on Mauna Kea, Hawaii.
Detailed descriptions of the design and performance of this
UCLA-built instrument are given elsewhere (McLean et al.\ 1998;
2000b). For this study, NIRSPEC was used in its cross-dispersed
echelle mode. High resolution spectra are dispersed across the
1024$\times$1024 InSb detector at 0$\farcs$143 per pixel while the
spatial scale in the cross-dispersed direction is 0$\farcs$19 per
pixel. An independent slit-viewing camera with a scale of 0$\farcs$18
per pixel is available for centering and guiding. With the
gratings used in NIRSPEC, the relationship between the blaze
wavelength ($\lambda_{b}$) and echelle order number (m) is
$m\lambda_{b}$ = 76.56 $\mu$m; together with the free spectral
range (see below), this equation gives the order location of a
given wavelength. The spectrometer was set up with the NIRSPEC-3
order-sorting filter and specific echelle and cross-dispersion
grating angles to record 11 echelle orders ($m$ = 66 to $m$ = 56)
covering the wavelength range from 1.15--1.36$\mu$m, corresponding
approximately to the standard $J$-band. The free spectral range
($\lambda_{b}/m$) at 1.255 $\mu$m (order 61) is $\sim$206~\AA, but
the effective dispersion is 0.179~\AA/pixel, allowing for only
183~\AA~ (89\%) of this order to be captured by the detector. In
fact, the captured wavelength range varies from 171~\AA~ (94\%) in
order 65 to 192 ~\AA~ (84\%) in order 58. Thus, because the
spectral interval captured by the detector is slightly smaller
than the free spectral range in each order, there are small gaps,
increasing with wavelength, in the total spectral coverage. Table
2 summarizes the spectral range for each order used in the
subsequent analysis. In practice, for an entrance slit 0$\farcs$43
(3 pixels) wide, the final spectral resolution in the reduced data
is $R \sim 20,000$, (or 15 km s$^{-1}$), compared to the
theoretical value of $R = 24,000$.  The average value of one
spectral resolution element is $\sim$0.625\AA~ (equivalent to 3.6
pixels) over most of the $J$-band region.

Spectroscopic observations were made as nodded pairs. Typically,
integrations of 600 s each were taken with the object placed at
two positions, designated A and B, separated by $\sim$7$''$ on the
$\sim$12$''$ long entrance slit of NIRSPEC. Shorter exposure times
were used for brighter objects. Exposures of 300 s per nod
position were used for 2MASS 0746+2000AB, Kelu-1AB and
2MASS 1507-1627, 120 s for Wolf 359 and 60 s per nod position for
G196-3A. Total integration times per object ranged from a few
minutes to 1.5 hours depending on the apparent $J$ magnitude.
Signal-to-noise ratios were typically greater than 20 (5\%) per
resolution element over most orders, and sometimes greater than
100 (1\% noise). Seeing conditions were $\sim0\farcs5-0\farcs6$
and therefore a slit width of 0$\farcs$43 (3 pixels) was used for
all observations, except in the case of 2MASS 1507$-$16, for which
we used a $0\farcs576$ (4 pixels) slit because of poorer seeing.
A0 V stars were observed at an airmass very close to that of the
target object to calibrate for absorption features caused by the
Earth's atmosphere. Arc lamp spectra, taken immediately after each
observation, and OH night sky lines in the observed spectra, were
used for wavelength and dispersion calibration. A white-light
spectrum and a corresponding dark frame were obtained for
flat-fielding.

\subsection{Data Reduction Methods}
For the data reduction we used REDSPEC, an IDL-based software
package developed at UCLA for NIRSPEC by S. Kim, L. Prato, and I.
McLean\footnote{See
http://www2.keck.hawaii.edu/inst/nirspec/redspec/index.html}. For
each echelle order, REDSPEC uses the position of the
two-dimensional spectra on the NIRSPEC array and the calibration
line spectra to construct spatial and spectral maps necessary to
transform the raw data onto a uniform grid. If the target spectrum
itself is too faint to provide the spatial rectification, then the
A0V star observed with the same set up was used instead. Although
four arc lamps are available, it is often the case that there are
too few well-distributed lines per echelle order for good spectral
rectification. Consequently, OH night sky lines were also used.
The dispersion was more than adequately fit by a second order
polynomial of the form $\lambda = c_{0} + c_{1} x + c_{2} x^{2}$
where $c_{1} \sim$ 0.17$\pm$0.01~\AA/pixel and $c_{2} \sim$ 7 x 10$^{-6}$
~\AA/pixel$^2$. To extract spectra free from
atmospheric background and uneven detector response, the
difference of an A/B image pair was formed and flat-fielded. The
flat-fielded difference frame was then rectified using the
spatial and spectral maps and the raw spectrum produced by summing
5$-$10 rows from each trace in the rectified image.  The extracted
traces (one positive, one negative) are subtracted again to produce a
positive spectrum with residual night sky emission line features
removed, unless a line was saturated.  In the $J$-band, none of the
night sky emission lines are saturated.  A0 V star spectra were reduced
in the same way, interpolating over the intrinsic Pa$\beta$
hydrogen absorption line at 1.28$\mu$m in the $J$-band spectra.
The raw target spectrum was then divided by the raw A0 V star
spectrum to remove telluric features.  The true slope of the target
spectrum was restored by multiplication with a blackbody spectrum
of T$_{eff}=9500$ K for an A0 V star (Tokunaga 2000). Finally, the
spectra reduced from multiple A/B pairs were averaged together to
improve the signal-to-noise ratio.

\section{J-band Spectral morphology at R$\sim$20,000}
\subsection{Overview}
For each of the 16 targets we have extracted 8 echelle orders (see
Table 2) yielding a total of 128 spectra. Before examining and
interpreting the new spectra in detail, it is very useful to have
a broad overview of the basic spectral features present and an
awareness of the general trends that occur in the high-resolution
$J$-band data as a function of spectral type. A convenient way of
doing this is to select a representative source for a few spectral
types and present all eight echelle orders on the same plot, thus
enabling the entire $J$-band to be viewed at a glance. Figures
1$-$6 show the reduced spectra of Wolf 359 (M6), 2MASS 0140+27
(M9), 2MASS 0345+25 (L0), 2MASS 1507$-$16 (L5), SDSS 0423$-$04AB
(T0) and 2MASS 0559$-$14 (T4.5). The double nature of SDSS
0423$-$04AB means that we do not have a true T0 spectrum, but lower
resolution studies (M03) show that J-band spectral variations are
relatively weak from L6 to T2 and therefore this binary remains a
useful proxy for a T0 dwarf. In these plots, echelle orders 58
through 65 are shown together; the remaining orders at the edges
of the $J$-band are too contaminated by strong atmospheric
absorption to be useful. For ease of comparison, all spectra are
shown in the laboratory reference frame and vacuum wavelengths are
used throughout; radial velocities and searches for radial
velocity variations will be reported and discussed in a separate
forthcoming paper (Prato et al. 2006 in prep.). Each order is normalized to
unity at the same wavelength. Comparison of the spectra in these
six figures, all of which have excellent signal to noise ratios
(at least 20:1 per pixel), shows that the region is densely
populated with numerous weak absorption features and a few
stronger lines. We will show that the fine-scale spectral
structure is real and repeatable, and that it is mainly attributable to FeH
or H$_{2}$O. The strongest atomic features are the doublets of K I
that occur in orders 61 and 65. These lines persist from M6-T4.5 but
clearly change their character with spectral type. In later
sections the K I lines will be singled out for closer inspection.
For reference, Table 3 summarizes the main spectral transitions
observed in the $J$-band over the spectral type range from M6-T4.5,
including the energy levels of the atomic transitions.

As shown in Figure 1, the M6 dwarf Wolf 359 has at least one
distinguishing feature in each order. Atomic lines of Al I at
1.31270 and 1.31543 $\mu$m appear in order 58. There is a
moderately strong line of Mn I at 1.29033 $\mu$m in order 59, plus
some weaker lines of Ti I at 1.28349 and 1.28505 $\mu$m. A
weak unresolved Na I doublet is seen at 1.26826 $\mu$m in order
60. The first pair of strong K I lines at 1.24357 and 1.25256
$\mu$m appears in order 61. Multiple weak absorption features
occur in both orders 62 and 63, the most notable grouping being
the set of lines around 1.222 $\mu$m. Several of the stronger
features have been identified with FeH from lower resolution
studies (Jones et al. 1996; Cushing et al.\ 2003; M03). Note
however, that a major FeH band head at 1.24 $\mu$m is just off the
detector at the short wavelength edge of this order. In order 64
there is a pair of strong Fe I lines at 1.18861 and 1.18873 $\mu$m
and another Fe I line at 1.19763 $\mu$m. Order 65 contains the
second set of strong K I lines, one at 1.169342 $\mu$m and the
close pair at 1.177286, 1.177606 $\mu$m.  In general, the lines
are relatively sharp and well-resolved. Wolf 359 is a bright
source and therefore the signal-to-noise ratio in this spectrum is
at least 100:1.

Following these spectral features order by order through Figures 1-6
reveals certain general trends as a function of spectral type.
Comparing the M9 object (Figure 2) with the M6 source (Figure 1) we see
that the Al I lines at 1.3127 and 1.3154 $\mu$m in order 58 are
somewhat weaker at M9 and then suddenly they are no longer present
at L0 (Figure 3), or in any later spectral types (Figures 4-6). This is
an important observation that relates to the M-L transition and we
will discuss the Al I lines in the next section. Throughout order
58 there are other weaker spectral features, the so-called
fine-scale spectral structure. This spectral structure becomes
more pronounced at M9 (Figure 2), seems broader in the L0 and L5
objects (Figures 3 and 4), weakens at T0 or more accurately from L6
to T2 (Figure 5) and then completely changes character by T4.5 (Figure
6). The most likely interpretation of this spectral sequence is
that it represents changes in the physical structure of these cool
atmospheres (temperature, pressure, chemistry). In subsequent
sections we compare opacity data for different molecular species
to identify the primary absorbers at each spectral class.

In order 59 the sharp Mn I line at 1.2903 $\mu$m seen at M6 and M9
(Figures 1 and 2) broadens and disappears after L0 (Figure 3). The
fine-scale spectral structure in this order is dominant until T0
composite type (Figure 5) when the spectrum becomes remarkably
smooth. Here again the high resolution data reveal a striking
effect, this time at the transition from L to T dwarfs. New
spectral structure develops in this order between types T0 at T4.5
(Figure 6) but, as was the case for order 58, the pattern is
different, indicating different atmospheric conditions.

The weak Na I line at 1.2683 $\mu$m detected in order 60 in the M6
object (Figure 1) is already absent in the M9 object (Figure 2).
Otherwise, the behavior of the fine-scale structure follows a
pattern similar to order 59 becoming remarkably weak at T0 (Figure
5) and leading to a smoother appearance for these spectra near the
L-T transition.

Order 61 contains one of the pairs of strong K I doublets located
at 1.2436 and 1.2525 $\mu$m. These lines deepen and widen slightly
from M6 to M9, and then become increasingly broader and shallower
from L0 to T4.5. NIRSPEC spectral order 61 also has many fine-scale
features attributable to molecular transitions. Two features, one
at 1.24637 and the other at 1.24825 $\mu$m have been identified
previously with FeH (Cushing et al. 2003). These FeH lines
strengthen slightly from M6 to M9 (Figures 1 and 2), become much
broader in the L dwarfs (Figures 3 and 4) and then vanish completely
in the T dwarfs (Figures 5 and 6) to leave, once again, a remarkably
smooth continuum between the K I lines.

Comparing orders 62 and 63 in Figures 1-6, the known FeH features in
these spectral bands strengthen from M6 to M9, broaden markedly at
L0 and remain strong and broad through L5 before becoming weaker
in the T0 and T4.5. As with the atomic lines, the individual FeH
lines seem to broaden significantly at the transition from
spectral types M9 to L0. Evidence of weak FeH absorption is still
present around 1.222 $\mu$m at spectral type T0, and possibly even
at T4.5, as shown in Figures 5 and 6, but this molecular species is
clearly not dominant in T dwarfs.

For the M6 dwarf (Figure 1), order 64 is characterized by a pair of
strong lines of Fe I at 1.18861 and 1.18873 $\mu$m that are easily
resolved, and another Fe I line at 1.19763 $\mu$m that is blended
with FeH. These features remain strong at M9 (Figure 2) and persist
into the L-dwarf range, becoming broader at L0 (Figure 3), and then
undetectable by L5 (Figure 4). From T0 to T4.5 (Figures 5 and 6), order
64 becomes increasingly chopped up by new spectral features, some
of which are quite sharp and deep. In section \S 3.3 we show that
these features are caused by absorption by \h2o.

Finally, there is order 65, which contains the second K I doublet
and exhibits some of the largest changes with spectral type. The
slightly weaker K I companion line at 1.17728 $\mu$m, only 3.3\AA~
from the longer wavelength member of the doublet is easily
resolved in the M9 object (Figure 2), already blended from line
broadening in the L0 (Figure 3), barely discernable at L5 (Figure 4),
and completely washed out by line broadening and numerous
molecular features at T4.5 (Figure 6). Order 65, being close to the
short wavelength edge of the $J$-band where terrestrial water
vapor absorption is expected, also contains many strong intrinsic
transitions of hot H$_{2}$O, for example, the feature at 1.175
$\mu$m.

\subsection{Al I, Fe I and Mn I; indicators of the M-L transition}
Figure 7 provides a more detailed view of the Al I doublet in
order 58. In this plot, the spectra for the M6, M9 and L0 objects
shown in Figures 1-3 are expanded and overlaid. Evidently, there is
significant spectral structure in this part of the spectrum making
it difficult to identify a true continuum level. All three
spectral types show consistent features, in particular the wide
depression containing the shorter wavelength Al I line. The
equivalent width of the Al I lines clearly decreases from M6 to
M9, but the change over these three spectral types is only about
25\%. Because this region of spectrum is contaminated by \h2o
absorption, it is difficult to obtain accurate equivalent widths.
A pseudo-equivalent width over a 4.2\AA~ interval centered on each
line was obtained relative to the local continuum in the troughs
where the Al I lines are found. For the stronger line of the pair
at 1.3127 $\mu$m, the measured values of equivalent width for the
M6 and M9 dwarfs respectively are 420$\pm$20 m\AA~ and 300$\pm$40
m\AA. At L0, however, the pseudo-equivalent width of this line is
$\le$40 m\AA. Clearly, at the transition from M9 to L0, both Al I
lines vanish completely. Although only three objects bridging this
transition were observed at high resolution, the conclusions given
here are supported fully by the results of our low resolution BD
spectroscopic survey (M03) where two objects of every spectral
type from M6 to L5 was included.

As shown in Table 3, these lines arise from absorption from an
energy level at 3.14 eV. Interestingly, the Na I line at 1.268
$\mu$m in order 60 is already absent in the M9, and careful
inspection shows that a somewhat broadened Mn I line at 1.290
$\mu$m in order 59 persists through L0. The Na I line is excited
from a high energy state at 3.6 eV whereas the Mn I line comes
from a state at only 2.1 eV. Thus, the sequence in which the lines
disappear is at least qualitatively consistent with thermal
excitation. But the abrupt loss of Al I lines at the classical M-L
boundary is too great to be explained by Boltzmann factors alone.
For example, for a temperature change from 2850 K from the M6 to
about 2400 K for the M9 say, the population of excited atoms in
the upper level would drop by 51\% and the equivalent width of the
line might change from 420 m\AA~ to about 210 m\AA~ if all other
factors remain the same. From M9 to L0 the change would be a
further 13\% assuming a change in effective temperature of 150 K.
Thus, the line should still be measurable with an equivalent width
of 140 $\pm$40 m\AA, or about one-third its value at M6. Yet, both
lines disappear abruptly. It is likely, from the models of Lodders
(2002), that aluminum has been sequestered in compounds such as
hibonite (CaAl$_{12}$O$_{19}$) and that this abrupt change in
absorption line strength is really caused by the sudden depletion
of aluminum as an absorber due to a significant change in
atmospheric chemistry, rather than simply a drop in effective
temperature. Gas temperatures typical of this transition are near
2000 K (Lodders 2002).

It is also curious that the intensity ratio of the components of
the Al I doublet is closer to 3:2 than the expected 2:1 ratio
based on their statistical weights. However, as shown in Figure 7,
this spectral region is highly complex with many overlapping
transitions which makes it difficult to determine the true
continuum level for each line. Alternatively, the peculiar line
ratios may be a non-LTE effect, or the result of line blending.

Another element that is also important for understanding the
temperature structure of these cool, dust-forming objects is iron.
As previously mentioned, order 64 contains a remarkably strong
pair of Fe I lines at 1.1886 and 1.1976 $\mu$m. The shorter
wavelength Fe I line is a resolved double with a separation of
1.2\AA~ in the M6 and M9 objects, but appears as a single broad
feature at L0. By L5 the Fe lines are completely absent.  Lower
resolution studies (M03) also suggested that Fe disappeared around
L2 or L3. These Fe I lines arise from low-lying energy levels near
2.2 eV, and gas phase iron requires temperatures above about 1700
K (Burrows et al. 2001). Combining the results that Al disappears
at L0 and Fe is no longer present by L3, and using the chemistry
temperature scale, suggests that there is about a 300 K temperature
change from L0 to L3, which is shallower but still consistent with
the interval of about 140 K per spectral type derived by Burgasser
et al. (2002) as well as the effective temperature scale of Golimowski
et al. (2004). As noted by Burgasser et al. (2002), temperatures
derived from condensation chemistry tend to be systematically
cooler by about 500 K than those derived from empirical
determinations of T$_{eff}$ using objects with known parallax.
These conclusions are not necessarily inconsistent if different
spectral features probe a range of optical depths in the
atmosphere.

Finally, we note the presence of several weak lines of
Ti I that arise from energy states near 1.4 eV, even lower than
those of the strong potassium lines. A strong Ti multiplet at 0.97 \um
has also been seen in the spectra of M dwarfs up to at least M9
(Cushing et al. 2005).  Unfortunately, these weak
lines are impossible to trace after M9.

\subsection{Fine-scale structure; the role of FeH and \h2o}
The astronomical $J$-band is bounded by \h2o absorption bands from
terrestrial water vapor. It is therefore no surprise that
high-temperature \h2o (hot steam) transitions intrinsic to M, L
and T dwarfs encroach far into the $J$-band from both the short
and long wavelength ends. These so-called infrared water bands are
difficult to model because millions of transitions are needed
(Partridge \& Schwenke 1997).  Typically, models over-estimate the
depth of the infrared water bands. In addition to \h2o, some of
the stronger non-atomic transition features are known to be
attributable to FeH from lower resolution studies (Jones et al.\
1996; McLean et al. 2000). These features occur at 1.2091, 1.2113,
1.2135 and 1.2221 $\mu$m. Cushing et al.\ (2003) verified the
features at 1.1939 and 1.2389 $\mu$m as the 0--1 and 1--2 band
heads of the F$^{4}\Delta$ -- X$^{4}\Delta$ system of FeH, and
attributed a blended feature described by McLean et al. (2000) at
1.2221 $\mu$m as the F$^{4}\Delta_{7/2}$ -- X$^{4}\Delta_{7/2}$
Q-branch. These authors also listed 24 other relatively strong
features lying within the $J$-band. In the Cushing et al. list, no
FeH features were identified in the wavelength interval covered by
our order 65, which includes the strong shorter-wavelength doublet
of K I, and only one feature (at 1.2464 $\mu$m) was tabulated for
order 61, where the other K I doublet dominates.

To identify many more of the complex fine-scale features seen in
the spectral sequences of Figures 1--6, we analyzed opacity
(cross-section) data for both FeH and H$_{2}$O, (R. Freedman 2003, private
communication) and utilized the FeH line list and transition
catalog by Phillips et al. (1987). We are also grateful to Adam
Burrows who provided CrH opacity data (Burrows et al. 2002) and
Linda Brown who provided new opacity calculations for \ch4 in the
$J$-band (L. Brown 2004, private communication).

Figure 8 is a detailed view of order 62 (1.221-1.239 $\mu$m) for
the M9 object 2MASS 0140+27. Superimposed on the M9 dwarf spectrum
is a normalized and scaled FeH opacity plot for a temperature of
2000 K and a pressure of 1 bar. We use this plot to assist in
identifying spectral features. This interesting region of the
$J$-band contains a feature which is seen in lower resolution
spectra as a broad, flat-bottomed line at 1.2221 $\mu$m (McLean et
al 2000a). Each transition in the opacity data can be correlated
with either 0--1 or 1--2 transitions of the F$^{4}\Delta$ --
X$^{4}\Delta$ system tabulated by Phillips et al. (1987). As shown
in Figure 8, the feature observed at 1.2221 $\mu$m and attributed to
the Q-branch by Cushing et al. (2003), is actually a composite of
four Q-branch and three P-branch transitions plus one R-branch
transition of FeH. Thus, at our higher resolution, the broad
flat-bottomed feature seen in lower resolution spectra is
completely resolved into eight separate transitions. These
transitions are the following: Q(0-1) at 1.22137 and 1.221383
$\mu$m blended, 1.221934, 1.222504 and 1.22305 $\mu$m; P(0-1) at
1.22166, 1.22219 and 1.22244 $\mu$m; R(0-1) at 1.22218 $\mu$m.
R-branch transitions tend to correspond closely to P-branch lines.
For example, the P(0-1) 1.22219 $\mu$m line is only 0.1\AA~ from
the R(0-1) transition just given. Also, the Q-branch line at
1.222504 $\mu$m is blended with the nearby P-branch transition at
1.22244 $\mu$m.

The remainder of order 62 is dominated by 5 R-branch lines and
about 30 P-branch transitions; all are members of the 0--1 band of
F$^{4}\Delta$ -- X$^{4}\Delta$ system. In this one order alone,
there are now more identified FeH transitions than previously
known for the entire $J$-band. Summing up across all NIRSPEC
orders, we identify over 200 matches to the FeH opacity data base
and the tables by Phillips et al. (1987). There is little doubt
that FeH is a dominant molecular absorber in late M and early L
spectral types.

Interestingly, there are several features that cannot be
identified with FeH transitions in the given opacity tables. For
example, a sharp line at 1.2227 $\mu$m contaminates the Q-branch
feature, and there are other isolated groups near 1.225, 1.228,
1.231 and 1.234 $\mu$m which also are not attributable to FeH.
Because these features have a dependence on spectral type that is
similar to FeH, they may be unknown FeH transitions or transitions
of CrH. To explore the latter possibility we also compared our
spectra to CrH opacity calculations by Burrows et al. (2002).
Although the CrH opacity data had a lower resolution than our
spectra, there was good coincidence for the features at 1.225,
1.228 and 1.231 $\mu$m, but the features near 1.234 $\mu$m could
not be identified with CrH. Although transitions of CrH
contaminate those of FeH, the strongest fine-scale features in
this part of the $J$-band are attributable to FeH rather than CrH.
Stronger CrH bands occur at shorter wavelengths (Kirkpatrick et
al. 1999).

We also investigated the fine-scale structure in orders 61 and 65
which contain the important K I doublets. Figure 9 provides a plot
of the FeH opacity (at 2000 K and 1 bar pressure) across order 61,
normalized for convenience, smoothed to match the resolution of
our spectra, and over-plotted with the spectra of our M9 and L5
dwarfs. Essentially every feature can be traced to FeH, with a few
notable exceptions. For example, absorption features at 1.2448 and
1.2458 $\mu$m are clearly associated with FeH, but two significant
absorptions bands in between these limits do not correlate with
FeH. Note also that the L5 spectral features are broader than the
equivalent features in the M9 object.

As shown in Figures 1-6, there is a general increase in the strength
of FeH from mid-M until about L4 and then a decrease in FeH
features towards later spectral types. Figure 10 shows a plot for
order 61 in which the L5 and the T4.5 dwarfs in our sample are
compared. In this figure H$_{2}$O opacity curves (at 1000 K and 1
bar) are over-plotted. The T4.5 spectrum lacks FeH, and its shallow
depressions and small features are consistent with the H$_{2}$O
opacity. The FeH feature near 1.245 $\mu$m appears strongest in
the early to mid Ls, while the \h2o transition at 1.1752 $\mu$m
seems to gain in strength systematically toward later spectral
types.

In order 65, no transitions for FeH are listed in the opacity
files at the shorter wavelengths. Therefore, in Figure 11, we plot
the H$_{2}$O opacity (at 1000 K and 1 bar) for order 65, and
over-plot with the spectrum of the L3.5 and T4.5 from our sample.
This pair of spectral types provides distinct morphological
samples. Most of the features in the T4.5 dwarf evidently correspond
to H$_{2}$O. There is a particularly strong \h2o feature at 1.1752
$\mu$m in the T4.5 dwarf, which is also present in the L3.5 object
at a weaker level. In fact, weak H$_{2}$O absorption is already
present even in late M objects.

We also looked for transitions associated with the $\phi$ bands of
TiO (Galehouse 1980; Phillips 1973). These bands have been
detected in young, low-gravity sub-stellar objects by McGovern et
al. (2004), but are not usually apparent in low resolution spectra
of older field dwarfs. A model spectrum kindly provided by D.
Saumon shows the location of numerous $\phi$-band transitions in
order 61, with noticeable band heads for the $\Delta\nu$=-1, (0-1)
and (1-2) transitions at 1.25644 and 1.24239 $\mu$m respectively.
Absorption features do occur at these wavelengths from M6-L5, but
these also coincide with FeH features and are less likely to be
TiO because none of the other TiO transitions of the $\phi$ band
are seen.

Finally, using new opacity data for \ch4 (L. Brown 2004, private
communication), we searched for evidence of methane transitions in
the T dwarfs. The strongest features should occur in order 65
which also contains the very broad shorter-wavelength K I doublet.
Unfortunately, this spectral region is already heavily blanketed
by \h2o absorption. Even in the high signal-to-noise spectra of
2MASS 0559-1404 (T4.5), there are no transitions that can be
uniquely attributed to \ch4.

\subsection{The potassium doublets}
Because the K I doublets dominate orders 61 and 65, and because
these strong lines clearly persist throughout almost the entire
M, L and T range, we plot 12 of the objects from Table 1 together in
Figures 12 and 13 to
provide a more continuous sequence for these features. For this
figure, the L4 GD165B has been left out because the L3.5
2MASS 0036+18, with better signal-to-noise ratio, has been
included; GD165B will be discussed separately. The other two
objects omitted from these plots are the peculiar L2, G196-3B, and
the peculiar T6 dwarf, 2MASS 0937+29; again, these objects are
discussed separately later. Each spectrum in Figures 12 and 13 has
been normalized to a continuum value of one at a common wavelength
and shifted for clarity by an additive constant along the y-axis.
The spectra have been ordered according to their published
classification (Optical types for L-dwarfs: Kirkpatrick et al. 1999,
2000, 2001; NIR for T dwarfs: Burgasser et al. 2006)
and aligned at the K~I rest wavelengths in vacuum.  As mentioned
already, radial velocity determinations will be presented in a
separate paper (Prato et al. 2006, in prep.).

As given in Table 3, the shorter wavelength K~I lines (in order
65) correspond to the multiplet designation 4p $^{2}$P$^{o}$$-$3d
$^{2}$D, and the order 61 pair to the 4p $^{2}$P$^{o}$$-$5s
$^{2}$S multiplet. All are transitions between states at
1.61$-$2.67 eV. The pair of K I lines in order 65 have almost
equal intensity at the line center, whereas in order 61, the
1.2436 $\mu$m line is always weaker than the 1.2526 $\mu$m line.
For both K I doublets, the line ratios are similar throughout the
spectral sequence.

One of the most prominent results, evident in both Figures 12 and
13, is the significant K~I line broadening in the later type
objects. Primarily because of its higher temperature, the M2.5 dwarf
manifests very narrow lines with almost no wings. There is a slight
contamination of the shortest wavelength K I line in order 65 from an
Fe I line for spectral types before L4. For the M2.5, M6 and M9
dwarfs, separate peaks are clearly discernable for the longer
wavelength line of the K I doublet in order 65 at 1.17761 $\mu$m and
the secondary K I line at 1.17728 $\mu$m, but for later spectral types
the 1.17761 $\mu$m lines are heavily blended and not discernable as
separate features. In the earlier type objects, the pair of lines near
1.1786 $\mu$m in order 65 are attributable to Ti and Fe.

\section{Analysis and Discussion}

\subsection{Correlating spectral features and spectral type}
To quantify the changes in the K I lines illustrated in Figures 12
and 13 as a function of spectral type we calculate three
quantities, a line depth, a line strength (equivalent width) and a
line width. Line depth is defined in terms of the measured flux
($F_{\lambda}$) at the line center compared to the average value
for the continuum. For convenience we construct the line depth as
$1-F_{line}/F_{cont}$ at the line center; a weak line has almost
the same flux in the line as in the continuum and its line depth
measure is therefore almost zero. Deep lines have a depth index
approaching unity. Determining the continuum or pseudo-continuum
level introduces the largest uncertainty into this ratio. The
continuum is estimated by fitting a sloping line across the
feature between the highest points on either side lying within
$\pm$ 50\AA~ of the line center. Multiple trials obtained by
varying the positions by a few Angstroms provide a mean value and
an estimate of the uncertainty from fluctuations in the
continuum level.

Equivalent width (in \AA) is also determined by fitting a
continuum line between two points on either side of the line,
summing the residual intensities and multiplying by the width of a
pixel. The same range and method is used as for the line depth.
Line width (in km s$^{-1}$) is defined as the full width of the
line at the intensity level halfway between the apparent local
continuum, defined above for the line depth and equivalent width,
and the minimum line intensity at the line center (FWHM).

Several sources contribute to the uncertainty in the derived
quantities. The continuum level is difficult to identify, photon
noise reduces the signal-to-noise ratio for the fainter sources,
and contamination by numerous weak features is dependent on
spectral type. Molecular line contamination results in additional
fluctuations in the measured K I quantities and gives larger
uncertainties, despite the good signal-to-noise ratio of most of
the spectra. Our final plotted uncertainties are based on
repetitive trial fits and photon noise estimates.

Table 4 and Table 5 provide the measured quantities for the 1.2525
and 1.2436 $\mu$m K I lines respectively. This pair of K I lines
(order 61) is preferred because the shorter wavelength doublet is
too heavily contaminated by H$_{2}$O absorption, and one of the K
I lines is itself a blend (\S 3). Figures 14, 15, and 16 show,
respectively, the line depth, equivalent width and velocity line
width as a function of spectral type for the 1.2525 $\mu$m line.

Several of the trends mentioned in the previous section are
confirmed. The K I line depth increases from a weak line in early
M dwarfs to a strong deep line in the early/mid L dwarfs before
decaying towards later spectral types. Interestingly, there is
considerable scatter in line depth among similar spectral types.
The equivalent width (designated by W) is better behaved.
Variations in defining a consistent continuum may have a large
effect on the line depth quantity, whereas the growth of the line
wings may overcome such variations when calculating W and FWHM.
Equivalent width increases to a broad maximum around L2-L4 and
then decreases again, remaining essentially constant from L5 -
T4.5. These results are consistent with our more extensive lower
resolution studies (M03). In contrast, the velocity line width of
the 1.2525 $\mu$m K I line (Figure 16) increases almost
monotonically and steeply with spectral type. The change in
velocity width is dramatic, ranging from $\sim$ 60 km s$^{-1}$ at
M6 to almost 500 km s$^{-1}$ at T4.5. This change does not result
from rotational broadening because the line develops extensive
wings characteristic of pressure broadening as discussed in the
next section. One L-dwarf appears to stand out from the others in
this sequence and is indicated by a star symbol. This object is
Kelu-1AB, and will be discussed separately in the next section.
The behavior of the 1.2436 $\mu$m K I line (Table 5) is similar.

The changing behavior of the K I lines can be explained as
follows. As the temperature decreases towards later spectral
types, the transition levels become less populated and the line
should become weaker. However, with decreasing temperature, dust
grains settle below the photosphere and the transparency of the
atmosphere at wavelengths where gas opacity is weak improves.
Thus, line formation can be observed at much greater depths and
pressures (Saumon et al. 2003). Higher pressures cause the
development of the broad line wings through collision broadening,
primarily with H$_{2}$ molecules (e.g. Burrows \& Voloyubev 2003),
and hence an increase in the FWHM of the lines. The greater column
depth of K I in these transparent regions also serves to increase
their equivalent widths. The behavior has been observed with the K
I and Na I fundamental doublets (Reid et al. 2000), and is one of
the unique properties of brown dwarf atmospheres. The pressure
broadening of the K I lines at J-band follow quantitative
expectations as well. The observed ratio between the K~I line
widths in the T4.5 and L5 spectra is about 2:1. As the T$_{eff}$
decreases, the pressure at $\tau$=2/3 increases as already
described. According to models of cloudy atmospheres, kindly
provided by D. Saumon (2003; private communication), the change in
the photospheric pressure at these wavelengths is a factor of 5
from 1600 K to 1000 K. These effective temperatures were chosen to
illustrate a prediction of the models over the typical range of
L/T spectral types.  This implies that the average kinetic energy
of the molecules of H$_{2}$ is five times greater, or that the
average speeds are $\sim$2.24 times larger. Pressure broadened
line widths should be proportional to the average velocity of the
disturbing atoms. Thus, the observed pressure broadening from L5
to T4.5 is consistent with the expected change.

In contrast to the K I lines, Figures 17 and 18 show the changing
trends with spectral type for two representative FeH and H$_{2}$O
features that fall within the orders containing the K I doublets;
data are provided in Table 6. For FeH we plot the line depth index
of the 1.245 $\mu$m feature in order 61, and for H$_{2}$O we use
the strong 1.175 $\mu$m feature in order 65. The FeH line depth
reaches a peak around L3-L4, whereas the H$_{2}$O line depth
increases more-or-less monotonically from M6 to T4.5. Once again
Kelu-1AB stands out, this time because the FeH bands are not as deep
as expected for an L2.

\subsection{Line shapes: rotation and pressure broadening effects}
In the previous section we found that the K~I line width (FWHM)
and the \h2o line depth provided the best correlations with
spectral type. The exception to the FWHM correlation is Kelu-1AB,
which {\it is} known to be a rapid rotator (Basri et al. 2000),
and which was recently discovered to be a $\sim$0\farcs3~ visual
binary (Liu \& Leggett 2005; Gelino et al. 2006), although the orbital velocity of 3 km
s$^{-1}$ is only a fraction of the $Vsin~i$. As will be shown
below, the K I lines are not really suitable for $Vsin~i$ studies.

Interestingly, Kelu-1AB is not abnormal in its \h2o ratio. The one
point in the \h2o plot of Figure 18 that does seem slightly high is
due in fact to GD165B (L4); this result is consistent with M03. As
the companion to a white dwarf, GD165B is most likely to be an old
L dwarf with a higher gravity. Burgasser et al. 2003 has shown
that late-type subdwarfs, which are also presumably older and have
higher surface gravities, can show stronger \h2o compared to
equivalently classified disk dwarfs.

The presence of a binary companion can impact the observed line
properties discussed above in several ways. For example, a
spectroscopic companion only partially resolved in velocity space
might produce a spectrum with anomalously broadened lines at
certain epochs. A spectroscopic companion of similar mass to the
primary object should not only manifest itself in broadened lines,
observed at a favorable phase, but also should almost double the
expected brightness of the system. Fainter companions, although
interesting in their own right, should not significantly impact
the primary dwarf spectrum if the mass ratio, $M_2/M_1$, is less
than $\sim$0.5 (Prato et al. 2002). Visual companions at
separations too great to noticeably effect the spectrum will also
cause the observed spectrum to appear brighter than expected for
the distance and age of the target system.  Four targets in Table
1 are visual binaries: 2MASS 0746+20 (L0.5), Kelu-1 (L2),
DENIS 0205$-$11 (L7) and SDSS 0423$-$04 (T0).  Signatures of velocity
variations in these spectra will be addressed in a forthcoming publication.

Spectral features in Kelu-1AB appear to be significantly broader
than other field dwarfs of similar temperature by a factor of 2-3,
consistent with its high rotational velocity of $60\pm5$ km/s
(Basri et al. 2000). In Figure 19 we compare the L2 dwarf, Kelu-1AB,
with the L0.5 and L3.5 dwarfs in our sample. Because no feature is
completely free of blending at this spectral resolution and
wavelength throughout the late M to T sequence, measuring precise
rotational velocities is challenging. Indeed, because of the
increasing complexity of molecular features with cooler effective
temperatures, most if not all of the lines in the mid-L to T
objects may be blends. However, the wealth of fine-scale spectral
structure attributable to FeH and \h2o throughout the $J$-band
does provide some possibility of detecting trends in $Vsin~i$,
especially because pressure broadening should not be a large
effect for molecular transitions.

Examining the rich FeH structure in order 61 (Figure 12), as well as
orders 62 and 63 (Figures 1-4) where there are no strong alkali
lines, it is clear that there is a sudden broadening of the FeH
lines at the transition from M9 to L0, and that all of the FeH
features remain broad after that transition. In fact, if the M9
spectra in orders 62 and 63 are smoothed by a factor of 3 using a
simple moving average, the resulting spectrum is a very good match
for the observed L0 spectrum. This sudden broadening could be
caused by either increased pressure broadening or an increase in
rotational velocities. Examination again of Figure 13 shows that
many of the \h2o spectral features that develop in the T dwarfs
are relatively sharp, indicating that they are not formed at the
same depth and pressure as the potassium line wings. Therefore, in
the coolest objects, we are seeing evidence of vertical structure
in the atmosphere. Like the FeH lines in the late M and early L
dwarfs, the \h2o features provide a better estimate of rotational
velocities than the alkali lines. We can de-convolve the
instrumental profile by using the observed widths of unresolved OH
night sky lines and the very narrow features seen in the M2.5
object. If we interpret the change in line width at the M9-L0
boundary as a rotational velocity, then after removal of the
instrumental profile, all of the early L dwarfs appear to be rapid
rotators with $Vsin~i$ $\ga$ 30 km s$^{-1}$.

Rotational velocities have been reported for some of the objects
in our target list. Basri et al. (2000) give $Vsin~i$ = 60$\pm$5
km s$^{-1}$ for Kelu-1AB and 22$\pm$5 km s$^{-1}$ for DENIS
0205$-$11AB and $\la$3 km s$^{-1}$ for Wolf 359 (Gl 406). The
L3.5 dwarf 2MASS 0036+18 is reported to have a $Vsin~i$ of
$\sim$15 km s$^{-1}$ by Schweitzer et al. (2001).  Several of our
targets were also observed by Zapatero Osorio et al. (2006) at
high resolution, enabling them to measure the $Vsin~i$ of 2MASS
0036+18, SDSS 1254-01, and 2MASS 0559-14.  We note with interest
that the rotational velocities they measure for these sources, as
well as for another dozen dwarfs, are all very close to 30 km
s$^{-1}$ within the uncertainties.  We will analyze in detail the
rotational velocities of our sample in a future paper; however, we
comment here that our Figures 12 and 13 suggest that the T0 and
T4.5 dwarfs in our sample, SDSS 0423-04 and 2MASS 0559-14, appear
to reflect larger values of $Vsin~i$.  If there is a tendency
towards more rapid rotation among the T dwarfs, it is not apparent
in the measurements of Zapatero Osorio et al. (2006).

Clarke et al. (2003) report time resolved spectroscopy of Kelu-1AB
used to search for variability in photospheric molecular species.
They confirm the short rotation period of the system and find
variable H$\alpha$ profiles.  No evidence for a spectroscopic
companion was detected and it appears to be a normal L dwarf apart
from the high rotation rate (Clarke et al. 2003). The recently
discovered 0.29\arcs binary contributes only about 3 km s$^{-1}$
in velocity shift and therefore has no impact on the measured
$Vsin~i$. However, it is interesting to ask whether or not both
components have the same $Vsin~i$. In the binary system Gl569B
(Zapatero Osorio et al. 2004), line broadening is 2-3 times
greater in one component than the other, although this has been
attributed to a nested spectroscopic binary (Simon et al. 2006). It is
possible that the higher rotation rate of Kelu-1AB is the result of
age. A younger L dwarf just past the accretion phase might have a
higher rotational velocity. On the other hand, as discussed in the
next section, the young L2 dwarf G196-3B does not have such a rapid
rotation, or is being viewed at a special angle.

\subsection{Surface gravity effects}
As shown by models (e.g. Burrows et al. 2001), the youngest brown
dwarfs are expected to be hotter and more luminous. As a
sub-stellar mass object cools over the first 100 Myr, its radius
will contract by a factor of 2-10, depending on initial mass, and
then remain almost constant at a value close to that of Jupiter's
radius. Consequently, a very young low-mass brown dwarf could be
observed with a much earlier spectral type than it will have when
older than 1 Gyr, and its surface gravity ($g = GM/R^{2}$) will be
less than that of an old brown dwarf with the same observed
spectral type. A lower surface gravity implies less pressure
broadening (P$\sim$g/k$_{r}$, where k$_{r}$ is the Rosseland mean
opacity) and therefore one expects the K I lines to be narrower in
such objects. The effect of surface gravity on the K I lines is
significant and has already been observed at lower resolution
(McGovern et al. 2004; Kirkpatrick et al. 2006). High resolution
infrared spectroscopy of brown dwarfs provides a means of measuring
surface gravities and hence estimating mass. Because of their greater
column depth the K I lines are much more sensitive to pressure and
therefore to surface gravity, whereas the weaker FeH and \h2o lines
formed high in the atmosphere provide a better probe of rotational broadening.

G196-3B, the companion to the M2.5 G196-3A, is classified as an L2
dwarf (Kirkpatrick et al. 1999). The G196-3AB system is believed
to be $\sim$20-300 Myr old, rather than the 1 Gyr thought to be
typical of field L dwarfs (Rebolo et al. 1998). Figure 20 shows
the order 61 spectrum of G196-3B plotted together with the L0.5
dwarf (2MASS 0746+20AB) and the L3.5 dwarf (2MASS 0036+18).
Clearly, although the signal-to-noise ratio is poorer, all of the
spectral features of G196-3B are much narrower than expected for
an L2 dwarf. Using high resolution far-red optical spectra, Basri
et al. (2000) measured 10 km s$^{-1}$ for the rotational velocity of
G196-3B. The infrared K~I lines have cores only half as wide as
those of the L0.5 and L3.5 dwarfs, and much less pronounced
pressure-broadened line wings. This behavior suggests that the
line profiles result primarily from lower surface gravity,
consistent with the age of this object.

\subsection{Metallicity effects}
Burgasser et al. (2002) noted that the T6 dwarf 2MASS 0937+29 is
peculiar because, despite having characteristics common to its
spectral class, it appears to have no K I lines in a
low-resolution $J$-band spectrum. If the K I lines were shallow
and very broad, or sharp and weak, then they might escape
detection at low spectral resolution. As shown in Figure 21
however, where we compare our high resolution spectrum of this T6
to that of the T4.5 dwarf 2MASS 0559$-$14, the K I lines are neither
weak nor broad, they are completely absent. Given the good
signal-to-noise ratio of this high resolution spectrum, there is no
possibility that the K I lines were simply too weak to
detect.

Disappearance of the K I lines is unexpected because other T6
dwarfs still show these features (Burgasser et al. 2002, M03). In fact, we have
observed the T5.5 2MASS 2356$-$15 which exhibits broad K I lines similar
to those of a T5. Strong \h2o absorption features are clearly
present in order 65 in the T6. These features show good
correlation with the same features in the T4.5 but there is clear
variation in the individual features. One possible reason for the
absence of the K I lines in 2MASS 0937+29 is that this object has a different
metallicity from other field dwarfs. Alternatively, the lines
could be veiled by dust, but the trend at these spectral types is
for dust to settle below the photosphere.  In any case, there are
other features in 2MASS 0937+29 that do not appear to be veiled.

2MASS 0937+29 was also classified as peculiar because of its
extremely blue NIR colors and because it also has a very red
optical spectrum for its type, as determined from CH$_{4}$ and
\h2o strengths. This combination of attributes led Burgasser et
al. (2002) to propose that 2MASS 0937+29 is an old, metal-poor
brown dwarf in which enhanced collision-induced H$_{2}$ absorption
in the $K$-band gives the unusual blue NIR color. If this is an
old T dwarf then it is also probably a fairly high-mass brown
dwarf that has cooled to the temperature of a T6, and consequently
it has a higher gravity, higher pressure atmosphere. The higher
gravity could also result in increased pressure broadening of the
\h2o lines in 2MASS 0937+29 which would explain why these
features seem muted and broader. In a recent study, Burgasser,
Burrows and Kirkpatrick (2006) have found that the high surface
gravity can result in a cooler effective temperature than
equivalent T6 dwarfs. It is known (e.g., M03) that K I lines
disappear around T7 or T8, hence it is possible that 2MASS
0937+29 is depleted in K I because of its low temperature rather
than a low metallicity. Because the high-resolution spectra
presented here show no residual trace of the K I lines, and
because the \h2o lines are unusually broad, we contend that
2MASS 0937+29 must be cooler and/or more metal poor than a normal T6 dwarf.\\

\section{Conclusions}
Using a sequence of high-resolution infrared $J$-band spectra
(R$\sim$20,000 or 15 km s$^{-1}$) obtained with NIRSPEC on the
Keck II telescope, we have studied the spectral morphology of
objects from M6 to T6. The principal results of the survey are as
follows:

(1) Hundreds of small-scale spectral features are identified to be
either FeH or H$_{2}$O absorption features. Over ten times as many
FeH features as previously identified in brown dwarf spectra are
now confirmed. A few features of CrH are also identified, but no
convincing transitions of TiO or \ch4 at J-band are found in this
sample. FeH features attain maximum strength in the mid-L dwarfs,
while H$_{2}$O absorption becomes steadily stronger towards later
spectral types.

(2) FeH and \h2o line widths are typically $\sim$20 km s$^{-1}$
for the late M dwarfs, but broaden abruptly by over a factor of
two at the M to L transition. We interpret this effect as evidence
for increased rotational velocities in L dwarfs.

(3) The doublet of Al I at 1.31270 and 1.31543 $\mu$m is shown to
be very sensitive to spectral type. This doublet weakens through
M9 and then vanishes abruptly between M9 and L0. We suggest that
this sudden disappearance is more a consequence of a transition in
atmospheric chemistry than a simple decrease in atomic population
levels resulting from a change in effective temperature. That is,
the aluminum atoms are suddenly bound up into molecules and
grains.

(4) The wings and line widths of the K I doublets at 1.16934,
1.17761 $\mu$m and at 1.24357, 1.25256 $\mu$m increase
systematically, while line depth weakens with later spectral type.
The equivalent width (W) of the K I features reaches a maximum in
the mid-L dwarfs, decreases and then remains almost constant
through T4.5. The K I line profiles begin to exhibit pressure
broadened wings as early as late M. Line widths (FWHM) range from
$\sim$50 km s$^{-1}$ at M5 to almost 500 km s$^{-1}$ at T4.5. This
effect is consistent with the much greater depth that is probed in
cool T dwarf atmospheres at $J$-band.

(5) As shown in Figure 12, a characteristic of the transition from
L to T dwarfs is the decay of FeH spectral structure, resulting in
a smooth spectrum at high resolution for late L dwarfs and early T
dwarfs, before \h2o dominant spectral structure develops.

(6) The young L2, G196-3B, exhibits very narrow K I lines without
extensive pressure-broadened wings, indicative of a lower gravity
atmosphere.

(7) Kelu-1AB, another L2, has exceptionally broad infrared lines,
including FeH and H$_{2}$O features, confirming its status as a
rapid rotator ($Vsin~i \sim$ 60 km s$^{-1}$).

(8) Finally, the peculiar T6 dwarf 2MASS 0937+29 displays a
complete absence of potassium, in contrast to other late T
objects. We interpret this as either a metallicity effect
(depletion of K atoms) or a cooler T$_{eff}$ for this high surface
gravity object.

Although the sample of objects of different spectral types is
relatively small, these high-resolution, high signal-to-noise
spectra of M, L and T dwarfs should provide an important benchmark
for the detailed development and improvement of model atmospheres.\\

\acknowledgements
The authors wish to thank the staff of the Keck Observatory for
their outstanding support. I.S.M. acknowledges the staff of the
UCLA Infrared Laboratory and colleagues James Graham (UCB), James
Larkin (UCLA) and Eric Becklin (UCLA) for their support throughout
the development of the NIRSPEC instrument. We thank Adam Burrows,
Katharina Lodders, Linda Brown, Didier Saumon, Richard Freedman,
Travis Barman and Mark Marley for helpful discussions and opacity
data.  Finally, we thank the anonymous referee for a careful and
complete critique of this paper.  Work by S.S.K. was supported by the Astrophysical
research Center for the Structure and Evolution of the Cosmos
(ARCSEC) of Korea Science and Engineering Foundation through the
Science Research Center (SRC) program. A.J.B. acknowledges support
by NASA through Hubble Fellowship grant HST-HF-01137.01 awarded by
the Space Telescope Science Institute, which is operated by the
Association of universities for research in Astronomy, Inc., for
NASA, under contract NAS 5-26555. This research has made use of the
NASA/IPAC Infrared Science Archive, which is operated by the Jet
Propulsion Laboratory, California Institute of Technology, under
contract with the National Aeronautics and Space Administration.  This
publication makes use of data from the Two Micron All Sky Survey,
which is a joint project of the University of Massachusetts and the
Infrared Processing and Analysis Center, funded by the National
Aeronautics and Space Administration and the National Science
Foundation. Finally, the authors wish to extend special thanks to
those of Hawaiian ancestry on whose sacred mountain we are privileged
to be guests.

\clearpage

\pagestyle{empty}

\begin{deluxetable}{lccccc}
\tablewidth{0pt} \tablecaption{\bf TARGET LIST, KEY PROPERTIES AND
OBSERVING LOG \label{tbl-1}} \tablehead{ \colhead{$ $} &
\colhead{R.A.} &
\colhead{Dec.} &
\colhead{Spectral} &
\colhead{$J$\tablenotemark{a}} &
\colhead{UT Date}\\
\colhead{Object} &
\colhead{(J2000.0)} &
\colhead{(J2000.0)} &
\colhead{Type\tablenotemark{b}} &
\colhead{(mag)} &
\colhead{of
Observation}}

\startdata
G196-3A            & 10 04 21.0 & 50 23 06 & M2.5 & 8.08$\pm$0.026 & 2002 Apr 23 \\
Wolf 359 (GJ 406)  & 10 56 28.9 & 07 00 53 & M6 & 7.09$\pm$0.024 & 2002 Apr 23 \\
2MASSW J0140026+270150  & 01 40 02.6 & 27 01 50 & M9 & 12.49$\pm$0.021 & 2000 Dec 4 \\
2MASP  J0345432+254023  & 03 45 43.2 & 25 40 23 & L0 & 14.00$\pm$0.027 & 2000 Dec 4 \\
2MASSI J0746425+200032\tablenotemark{c}  & 07 46 42.6 & 20 00 32 & L0.5 & 11.76$\pm$0.020 & 2002 Jan 1\\
Kelu-1\tablenotemark{c} & 13 05 40.2 &$-$25 41 06 & L2 & 13.41$\pm$0.026 & 2003 May 12\\
G196-3B            & 10 04 21.0 & 50 23 06 & L2 & 14.83$\pm$0.047 & 2002 Apr 23\\
2MASSW J0036159+182110  & 00 36 16.2 & 18 21 10 & L3.5 & 12.47$\pm$0.027 & 2000 Dec 4\\
GD165B             & 14 24 39.1 & 09 17 10 & L4  & 15.69$\pm$0.078 & 2003 May 13\\
2MASSW J1507476$-$162738  & 15 07 47.7 &$-$16 27 39 & L5 & 12.83$\pm$0.027 & 2000 Apr 25\\
DENIS-P J0205.4$-$1159\tablenotemark{c} & 02 05 29.4 &$-$11 59 30 & L7 & 14.59$\pm$0.030 & 2001 Oct 9\\
SDSSp J042348.57$-$041403.5\tablenotemark{c}   & 04 23 48.6 &$-$04 14 04 & T0  & 14.47$\pm$0.027  & 2001 Oct 9\\
SDSSp J125453.90$-$012247.4  & 12 54 53.9 &$-$01 22 47 & T2  & 14.89$\pm$0.035 & 2003 May 14\\
2MASS J05591914$-$1404488  & 05 59 19.1 &$-$14 04 48 & T4.5 & 13.80$\pm$0.024  & 2001 Oct 9\\
2MASSI J2356547$-$155310  & 23 56 54.8 &$-$15 53 11 & T5.5 & 15.82$\pm$0.057 & 2005 July 19\\
2MASSI J0937347+293142 & 09 37 34.7 & 29 31 42 & T6p & 14.65$\pm$0.036 & 2003 May 12\\
\enddata
\tablenotetext{a}{From 2MASS All-Sky Point Source Catalog}
\tablenotetext{b}{Spectral types for L dwarfs are optical from Kirkpatrick et
  al. (1999, 2000, 2001);
T dwarf spectral types are near-infrared from Burgasser et al. (2006).}\\
\tablenotetext{c}{Known double or multiple object.}
\end{deluxetable}

\clearpage
\pagestyle{plaintop}

\begin{deluxetable}{lcccc}
\tablewidth{0pt} \tablecaption{\bf WAVELENGTH AND DISPERSION
PROPERTIES OF THE $J$-BAND\tablenotemark{a} NIRSPEC ECHELLE ORDERS \label{tbl-2}}
\tablehead{\colhead{$ $} &
\colhead{Wavelengths} &
\colhead{Interval} &
\colhead{Dispersion} &
\colhead{\% Free}\\
\colhead{Order} &
\colhead{(microns)} &
\colhead{(\AA)} &
\colhead{(\AA~per pixel)} &
\colhead{Spectral Range}}

\startdata
58 & 1.30447--1.32370 & 192.3 & 0.188 & 84.5 \\
59 & 1.28262--1.30151 & 188.9 & 0.184 & 85.9 \\
60 & 1.26137--1.27999 & 186.2 & 0.182 & 87.5 \\
61 & 1.24081--1.25913 & 183.2 & 0.179 & 89.0 \\
62 & 1.22093--1.23899 & 180.6 & 0.176 & 90.7 \\
63 & 1.20168--1.21938 & 177.0 & 0.173 & 91.7 \\
64 & 1.18293--1.20011 & 171.8 & 0.168 & 91.9 \\
65 & 1.16496--1.18207 & 171.1 & 0.167 & 94.4 \\
\enddata
\tablenotetext{a}{NIRSPEC-3 blocking filter ($\lambda\lambda$
1.143-1.375 $\mu$m); echelle angle = 63.00 degrees; grating angle
= 34.08 degrees.}\\
\end{deluxetable}

\clearpage

\begin{deluxetable}{lccc}
\tablewidth{0pt} \tablecaption{\bf SPECTRAL FEATURES IN THE
$J$-BAND NIRSPEC ECHELLE ORDERS \label{tbl-3}}
\tablehead{
\colhead{$ $} &
\colhead{Wavelength\tablenotemark{a}} &
\colhead{$ $} &
\colhead{Level Energy} \\
\colhead{Feature} &
\colhead{($\mu$m)} &
\colhead{Transition} &
\colhead{(eV)} }
\startdata
Al I & 1.3127007 & 4s $^{2}$S$_{1/2}$ - 4p $^{2}$P$_{3/2}$ & 3.143-4.087 \\
Al I & 1.3154345 & 4s $^{2}$S$_{1/2}$ - 4p $^{2}$P$_{1/2}$ & 3.143-4.085 \\
CrH\tablenotemark{b} & 1.18? & 6 bands of A$^{6}$ $\Sigma^{+}$ - X$^{6}$ $\Sigma^{+}$ & \nodata \\
Fe I & 1.1693174 & a$^{5}$P$_{1}$ - z$^{5}$D$^{o}$$_{1}$   & 2.223-3.283 \\
Fe I & 1.1786490 & b$^{3}$P$_{2}$ - z$^{3}$D$^{o}$$_{3}$   & 2.831-3.884 \\
Fe I & 1.1886098 & a$^{5}$P$_{2}$ - z$^{5}$D$^{o}$$_{3}$   & 2.198-3.241 \\
Fe I & 1.1887337 & a$^{5}$P$_{1}$ - z$^{5}$D$^{o}$$_{2}$   & 2.223-3.266 \\
Fe I & 1.1976325 & a$^{5}$P$_{3}$ - z$^{5}$D$^{o}$$_{4}$   & 2.176-3.211 \\
FeH\tablenotemark{c}  & 1.1939 band head & 0-1 band of F$^{4}$ $\Delta$-X$^{4}$ $\Delta$ & \nodata \\
FeH\tablenotemark{c}  & 1.2389 band head & 1-2 band of F$^{4}$ $\Delta$-X$^{4}$ $\Delta$ & \nodata \\
H$_{2}$O\tablenotemark{d} & 1.135 & $\nu_{1}+\nu_{2}+\nu_{3}$ & \nodata \\
H$_{2}$O\tablenotemark{d} & 1.331 & 2$\nu_{3}$ band & \nodata \\
K I  & 1.1693420 & 4p $^{2}$P$_{1/2}$ - 3d $^{2}$D$_{3/2}$ & 1.610-2.670 \\
K I  & 1.1772861 & 4p $^{2}$P$_{3/2}$ - 3d $^{2}$D$_{3/2}$ & 1.617-2.670 \\
K I  & 1.1776061 & 4p $^{2}$P$_{3/2}$ - 3d $^{2}$D$_{5/2}$ & 1.617-2.670 \\
K I  & 1.2435675 & 4p $^{2}$P$_{1/2}$ - 5s $^{2}$S$_{1/2}$ & 1.610-2.607 \\
K I  & 1.2525560 & 4p $^{2}$P$_{3/2}$ - 5s $^{2}$S$_{1/2}$ & 1.617-2.607 \\
Mn I & 1.290329  & a$^{6}$D$_{9/2}$ - z$^{6}$P$_{7/2}$     & 2.114-3.075 \\
Na I\tablenotemark{e} & 1.268261  & 3d $^{2}$D$_{5/2,7/2}$ - 5f $^{2}$F$_{5/2}$ & 3.617-4.595 \\
Na I\tablenotemark{e} & 1.268269  & 3d $^{2}$D$_{3/2}$ - 5f $^{2}$F$_{5/2}$ & 3.617-4.595 \\
Ti I & 1.1896124 & b$^{3}$F$_{2}$ - z$^{3}$D$^{o}$$_{1}$   & 1.430-2.472 \\
Ti I & 1.2674567 & b$^{3}$F$_{2}$ - z$^{3}$F$^{o}$$_{3}$   & 1.430-2.408 \\
Ti I & 1.2834947 & b$^{3}$F$_{2}$ - z$^{3}$F$^{o}$$_{2}$   & 1.430-2.396 \\
Ti I & 1.2850544 & b$^{3}$F$_{3}$ - z$^{3}$F$^{o}$$_{3}$   & 1.443-2.408 \\
\enddata
\tablenotetext{a}{All wavelengths are vacuum; atomic wavelengths are from NIST. }
\tablenotetext{b}{Burrows et al. 2002}
\tablenotetext{c}{Phillips et al. 1987}
\tablenotetext{d}{Auman et al. 1967}
\tablenotetext{e}{Blended}
\end{deluxetable}

\clearpage

\pagestyle{empty}

\begin{deluxetable}{llccc}
\tablewidth{0pt} \tablecaption{\bf 1.2525 $\mu$m KI LINE
PROPERTIES \label{tbl-4}} \tablehead{ \colhead{Spectral} &
\colhead{$ $} & \colhead{Line} & \colhead{W} &
\colhead{FWHM} \\
\colhead{Type} &
\colhead{Object} &
\colhead{Depth} &
\colhead{(\AA)} &
\colhead{(km s$^{-1}$)} }

\startdata
M2.5 & G196-3A            & 0.36$\pm$0.03 &  1.3$\pm$0.4 & 39$\pm$4   \\
M6   & Wolf 359           & 0.72$\pm$0.01 &  5.2$\pm$0.5 & 64$\pm$6   \\
M9   & 2MASS J0140+27  & 0.80$\pm$0.01 &  7.5$\pm$0.7 & 110$\pm$11 \\
L0   & 2MASS J0345+25  & 0.66$\pm$0.02 &  9.3$\pm$0.9 & 180$\pm$18 \\
L0.5 & 2MASS J0746+20AB  & 0.71$\pm$0.01 & 11.5$\pm$1.1 & 210$\pm$21 \\
L2   & Kelu-1AB             & 0.60$\pm$0.02 & 14.1$\pm$1.4 & 320$\pm$32 \\
L3.5 & 2MASS J0036+18  & 0.69$\pm$0.02 & 14.4$\pm$1.4 & 240$\pm$24 \\
L4   & GD165B             & 0.80$\pm$0.01 & 12.6$\pm$1.3 & 230$\pm$23 \\
L5   & 2MASS J1507$-$16  & 0.68$\pm$0.02 & 10.0$\pm$1.0 & 240$\pm$24 \\
L7   & DENIS J0205$-$11AB & 0.47$\pm$0.05 &  8.6$\pm$1.3 & 290$\pm$29 \\
T0   & SDSS J0423$-$04AB   & 0.47$\pm$0.06 &  9.3$\pm$1.4 & 240$\pm$24 \\
T2   & SDSS J1254$-$01   & 0.60$\pm$0.05 &  9.8$\pm$1.5 & 270$\pm$27 \\
T4.5 & 2MASS J0559$-$14  & 0.41$\pm$0.07 &  9.4$\pm$1.4 & 490$\pm$50 \\
\enddata
\end{deluxetable}

\clearpage

\pagestyle{empty}

\begin{deluxetable}{llccc}
\tablewidth{0pt} \tablecaption{\bf 1.243 $\mu$m KI LINE PROPERTIES
\label{tbl-5}} \tablehead{ \colhead{Spectral} & \colhead{$ $} &
\colhead{Line} & \colhead{W} &
\colhead{FWHM} \\
\colhead{Type} &
\colhead{Object} &
\colhead{Depth} &
\colhead{(\AA)} &
\colhead{(km s$^{-1}$)} }

\startdata
M2.5 & G196-3A            & 0.24$\pm$0.04 &  1.3$\pm$0.1 & 47$\pm$5   \\
M6   & Wolf 359           & 0.65$\pm$0.02 &  5.6$\pm$0.6 & 65$\pm$7   \\
M9   & 2MASS J0140+27  & 0.77$\pm$0.01 &  9.0$\pm$0.9 & 78$\pm$8 \\
L0   & 2MASS J0345+25  & 0.61$\pm$0.02 & 11.5$\pm$1.2 & 220$\pm$22 \\
L0.5 & 2MASS J0746+20AB  & 0.67$\pm$0.02 & 14.1$\pm$1.4 & 230$\pm$23 \\
L2   & Kelu-1AB             & 0.57$\pm$0.02 & 14.1$\pm$1.4 & 320$\pm$31 \\
L3.5 & 2MASS J0036+18  & 0.63$\pm$0.02 & 11.4$\pm$1.1 & 290$\pm$29 \\
L4   & GD165B             & 0.75$\pm$0.01 & 14.0$\pm$1.4 & 150$\pm$15 \\
L5   & 2MASS J1507$-$16  & 0.62$\pm$0.02 & 14.7$\pm$1.5 & 270$\pm$27 \\
L7   & DENIS J0205$-$11AB & 0.36$\pm$0.06 &  8.2$\pm$1.2 & 390$\pm$39 \\
T0   & SDSS J0423$-$04AB   & 0.36$\pm$0.08 &  7.5$\pm$1.1 & 300$\pm$30 \\
T2   & SDSS J1254$-$01   & 0.49$\pm$0.06 &  7.9$\pm$1.2 & 260$\pm$26 \\
T4.5 & 2MASS J0559$-$14  & 0.29$\pm$0.09 &  9.4$\pm$1.4 & 460$\pm$46 \\
\enddata
\end{deluxetable}

\clearpage

\pagestyle{empty}

\begin{deluxetable}{llcc}
\tablewidth{0pt} \tablecaption{\bf PROPERTIES OF THE 1.245 $\mu$m
FeH AND 1.1752 $\mu$m \h2o FEATURES \label{tbl-6}}
\tablehead{
\colhead{Spectral} &
\colhead{$ $} &
\colhead{FeH} &
\colhead{\h2o} \\
\colhead{Type} &
\colhead{Object} &
\colhead{Depth} &
\colhead{Depth}}

\startdata
M2.5 & G196-3A            & 0.03$\pm$0.04 &  0.022$\pm$0.01 \\
M6   & Wolf 359           & 0.12$\pm$0.04 &  0.080$\pm$0.016 \\
M9   & 2MASS J0140+27  & 0.17$\pm$0.04 &  0.140$\pm$0.029 \\
L0   & 2MASS J0345+25  & 0.25$\pm$0.04 &  0.110$\pm$0.022 \\
L0.5 & 2MASS J0746+20AB  & 0.29$\pm$0.04 &  0.100$\pm$0.02 \\
L2   & Kelu-1AB             & 0.15$\pm$0.04 &  0.110$\pm$0.022 \\
L3.5 & 2MASS J0036+18  & 0.28$\pm$0.04 &  0.160$\pm$0.023 \\
L4   & GD165B             & 0.35$\pm$0.05 &  0.35 $\pm$0.05 \\
L5   & 2MASS J1507$-$16  & 0.23$\pm$0.04 &  0.220$\pm$0.044 \\
L7   & DENIS J0205$-$11AB & 0.19$\pm$0.08 &  0.270$\pm$0.067 \\
T0   & SDSS J0423$-$04AB   & 0.08$\pm$0.07 &  0.280$\pm$0.055 \\
T2   & SDSS J1254$-$01   & 0.16$\pm$0.07 &  0.380$\pm$0.075 \\
T4.5 & 2MASS J0559$-$14  & 0.06$\pm$0.09 &  0.620$\pm$0.093 \\
\enddata
\end{deluxetable}

\clearpage
\thispagestyle{empty}

\begin{figure}[!htp]
\vspace*{-10mm}
\epsscale{0.95} \plotone{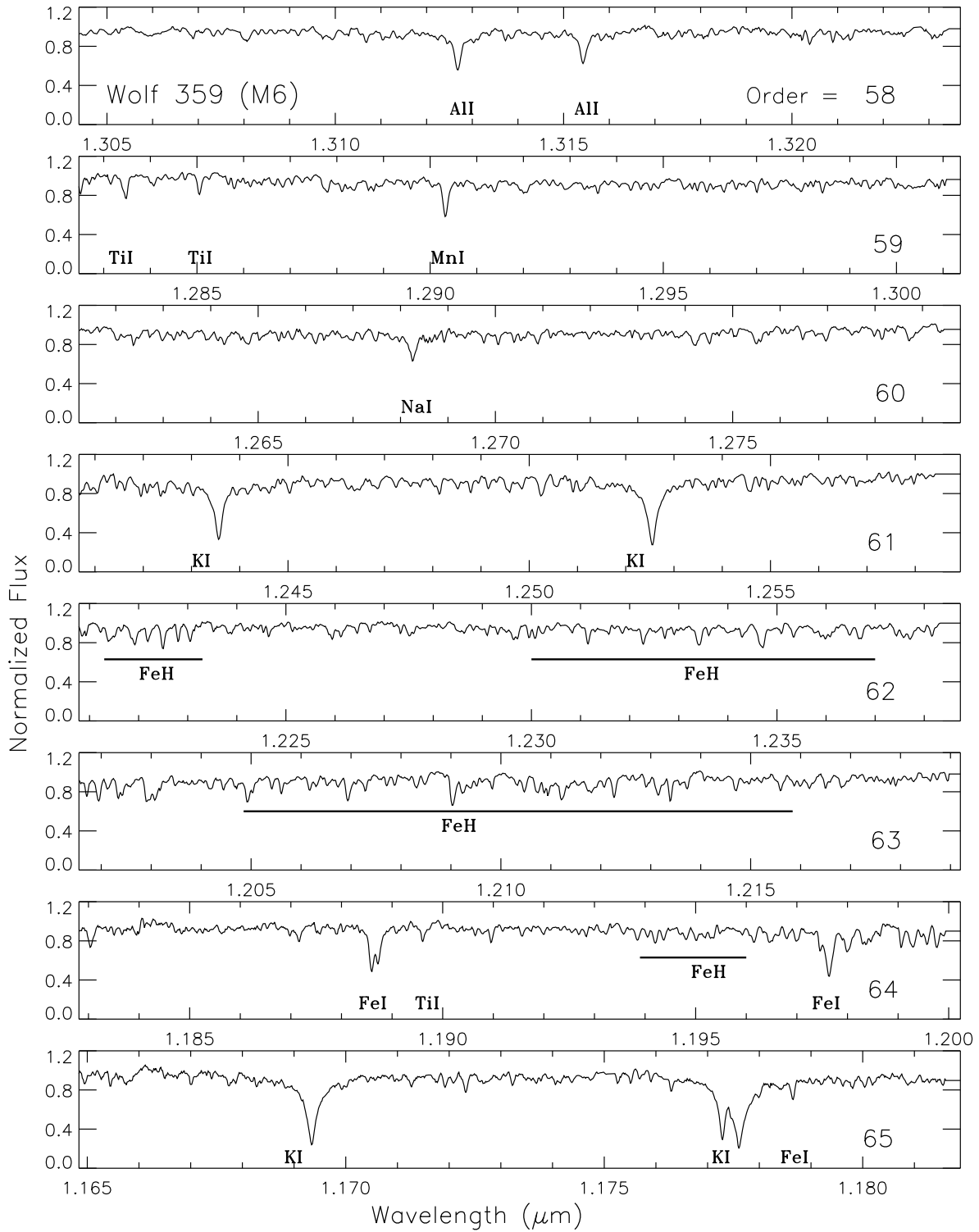} \caption{Eight NIRSPEC echelle
orders covering the wavelength range 1.165-1.323 $\mu$m are shown
for the M6 dwarf Wolf 359 (GJ~406). The resolving power is $\sim$
20,000 (15 km s$^{-1}$) giving a resolution of 0.625\AA~ at 1.250
$\mu$m. One pixel is $\sim$0.17\AA~ and wavelength tick marks are
10\AA~ apart. Spectra are normalized to unity within each order,
and shifted to the laboratory vacuum wavelengths. Key spectral
features are labeled. The signal-to-noise ratio is close to
100:1. Essentially all of the fine spectral structure is real and
repeatable; see text for details.}
\end{figure}
\epsscale{1}

\clearpage
\pagestyle{plaintop}
\begin{figure}[!htp]
\epsscale{0.95} \plotone{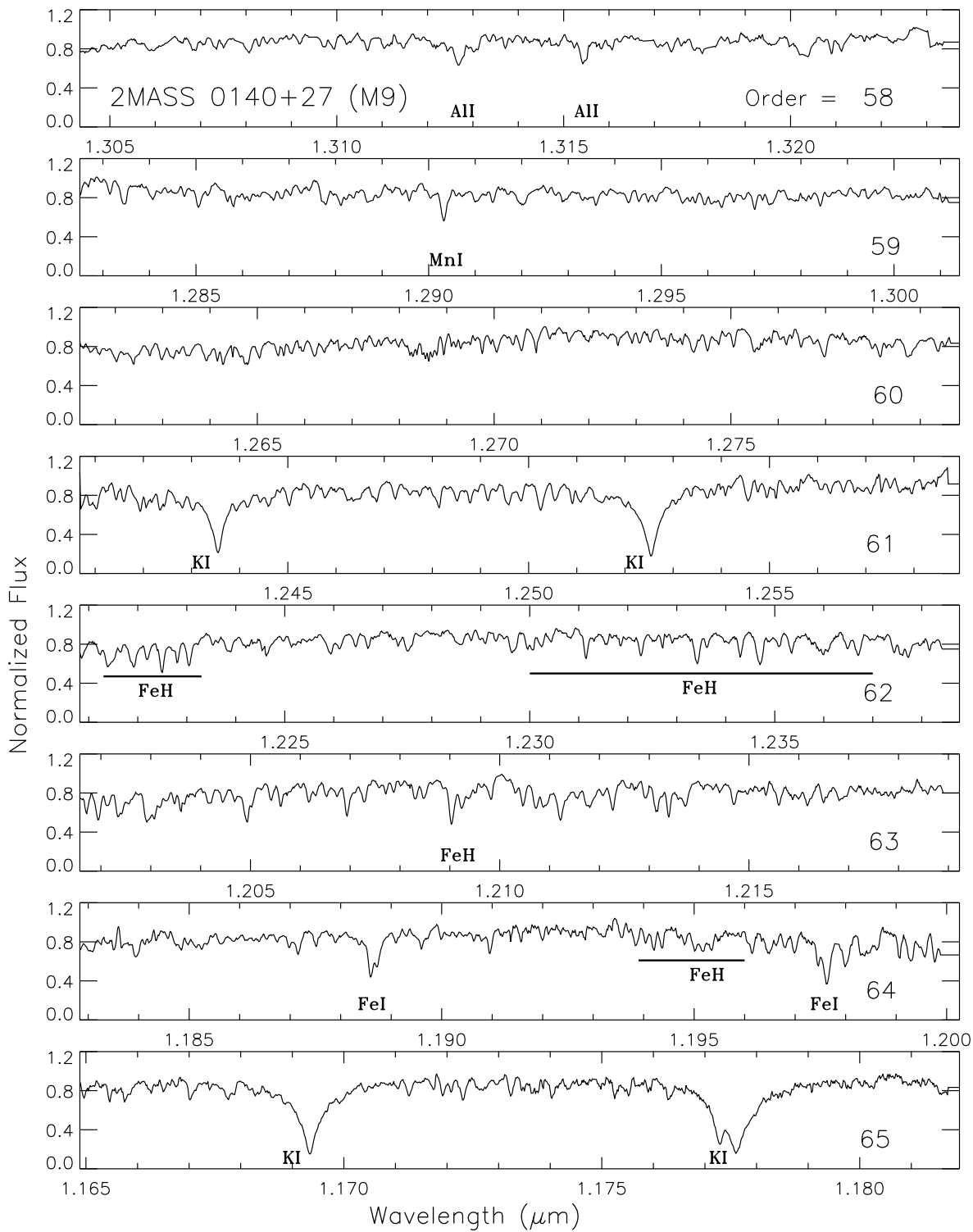} \caption{Eight NIRSPEC echelle
orders covering the wavelength range 1.165-1.323 $\mu$m are shown
for the M9 dwarf 2MASS 0140+27. As for Figure 1, the resolving
power is $\sim$ 20,000, the spectra are normalized to unity within
each order and the spectra have been shifted to the laboratory
vacuum wavelength scale. Key spectral features are labeled. The
noise level ($\la$2\%) is almost imperceptible and the fine
spectral structure is real and repeatable.}
\end{figure}
\epsscale{1}

\clearpage

\begin{figure}[!htp]
\epsscale{0.95} \plotone{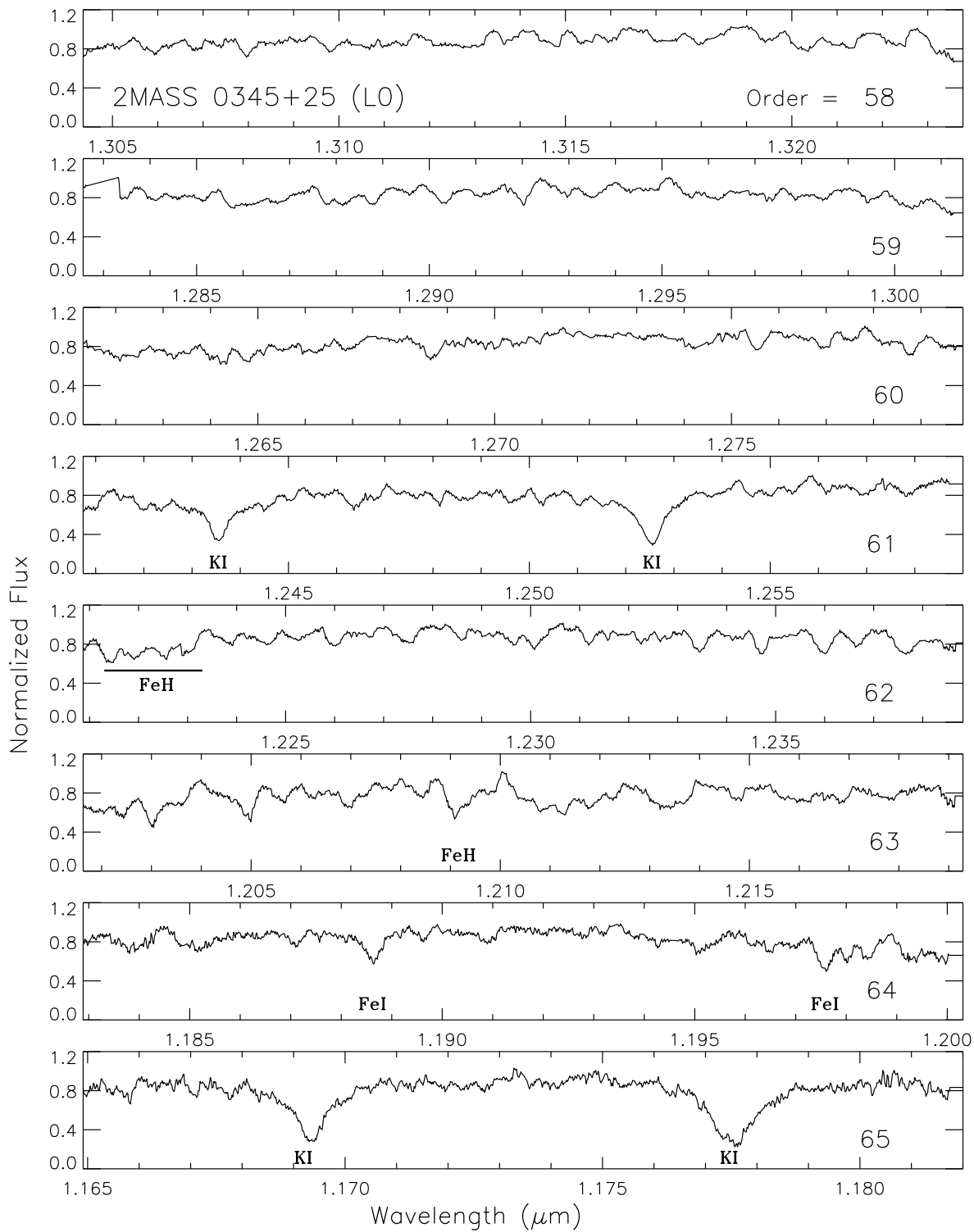} \caption{As for Figures 1 and 2,
eight orders covering the wavelength range 1.165-1.323 $\mu$m are
shown for the L0 dwarf 2MASS 0345+25. Spectra are normalized
to unity within each order and shifted to laboratory vacuum
wavelengths. Note the significant change in line widths; all
features are broad. Key spectral features are labeled.}
\end{figure}
\epsscale{1}

\clearpage

\begin{figure}[!htp]
\epsscale{0.95} \plotone{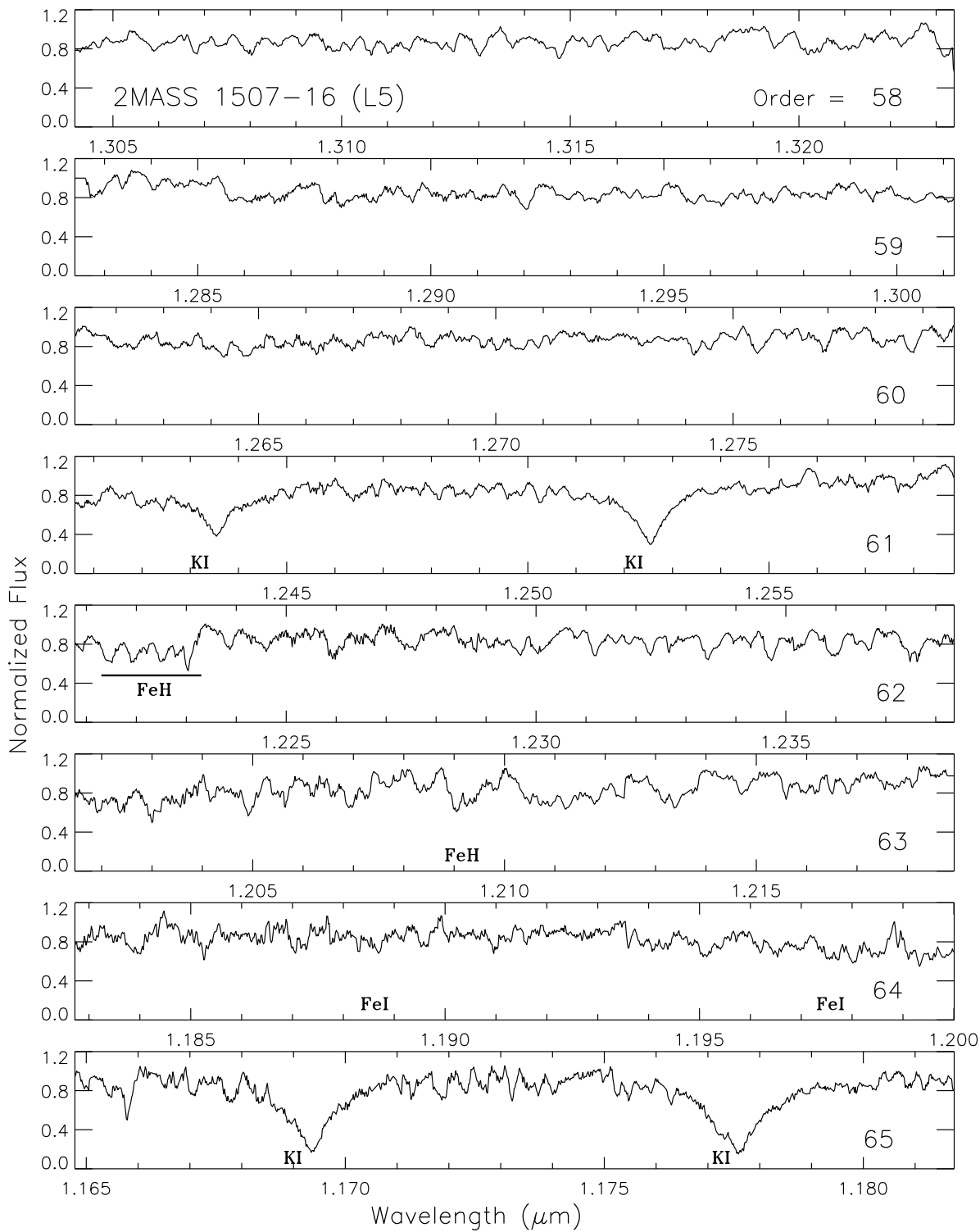} \caption{As for Figure 1-3, eight
orders covering the wavelength range 1.165-1.323 $\mu$m are shown
for the L5 dwarf 2MASS 1507$-$16. This spectrum is at a slightly lower
resolving power than the other objects because a wider slit was used
(see \S 2.1). Spectra are normalized to
unity within each order and shifted to laboratory vacuum
wavelengths. Key spectral features are labeled. The
signal-to-noise ratio is better than 30:1 over most of the
spectrum.}
\end{figure}
\epsscale{1}

\clearpage

\begin{figure}[!htp]
\epsscale{0.95} \plotone{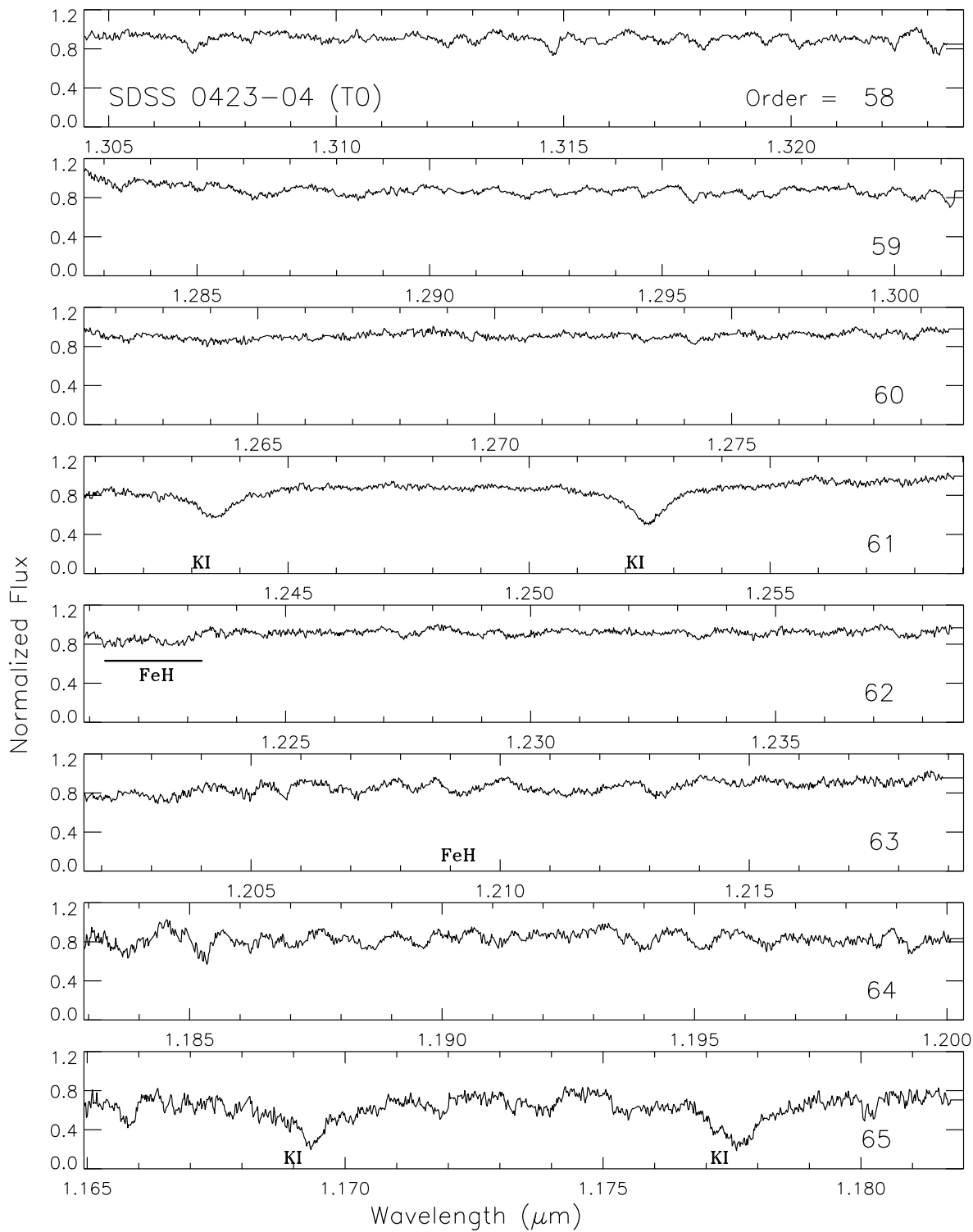} \caption{The same eight NIRSPEC
echelle orders covering the wavelength range 1.165-1.323 $\mu$m as
shown in Figures 1-4 are repeated here for the T0 dwarf SDSS
0423$-$04AB. The resolving power is $\sim$ 20,000, spectra are
normalized to unity within each order and vacuum laboratory
wavelengths are used. Note how smooth the spectra have become
relative to earlier spectral types. Key spectral features are
labeled. The noise level is about 3-5\% of the continuum.}
\end{figure}
\epsscale{1}

\clearpage

\begin{figure}[!htp]
\epsscale{0.95} \plotone{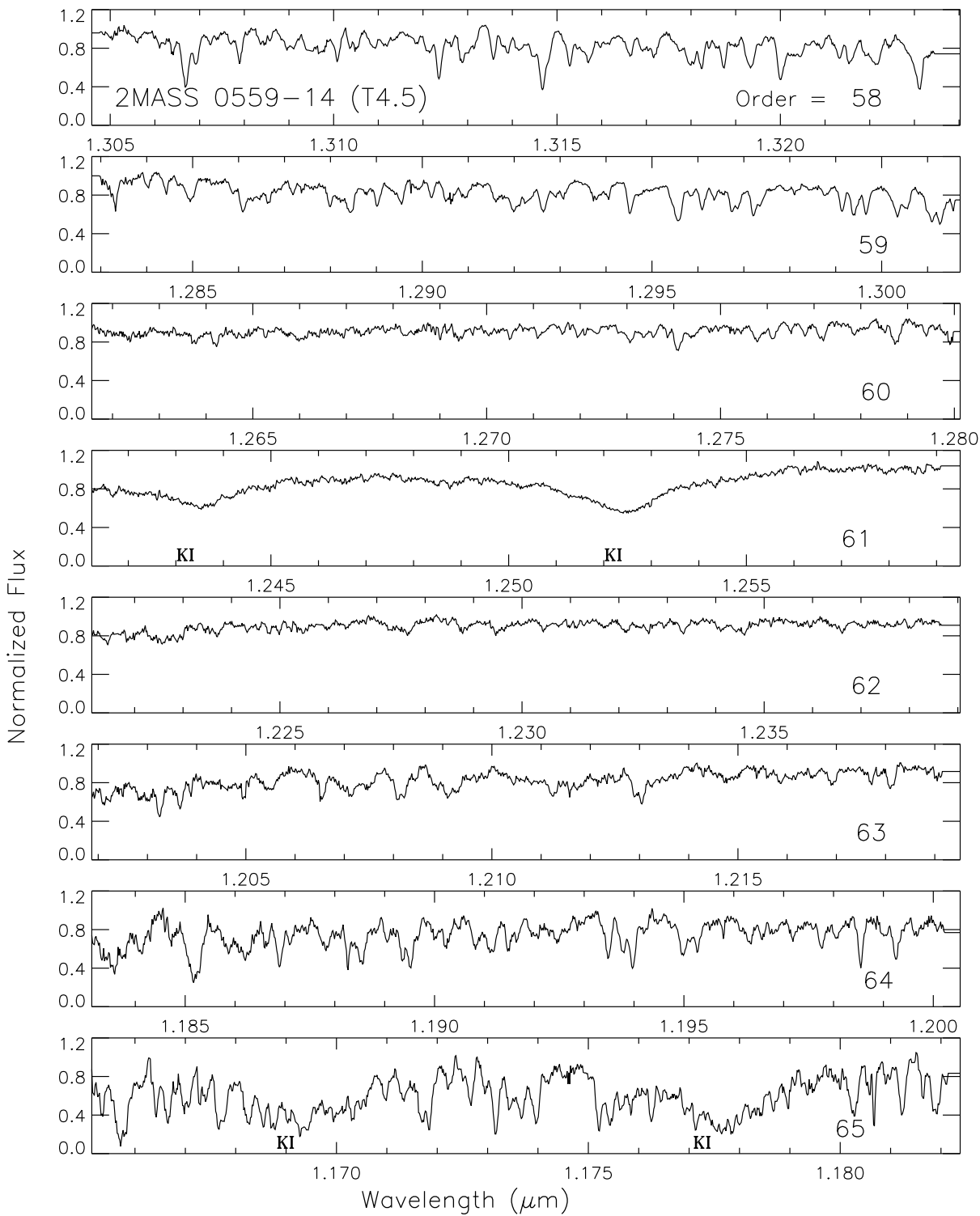} \caption{Eight orders covering
the wavelength range 1.165-1.323 $\mu$m are shown for the T4.5 dwarf
2MASS 0559$-$14. As in the previous figures, the spectra have
the same resolution, are normalized to unity within each order and
shifted to vacuum laboratory wavelengths. Note the increase in
absorption features, of \h2o, at both ends of the range. The
noise level ($\la$2\%) is indicated by the smoothness in order
61.}
\end{figure}
\epsscale{1}

\clearpage

\begin{figure}[!htp]
\includegraphics[scale=0.65,angle=90]{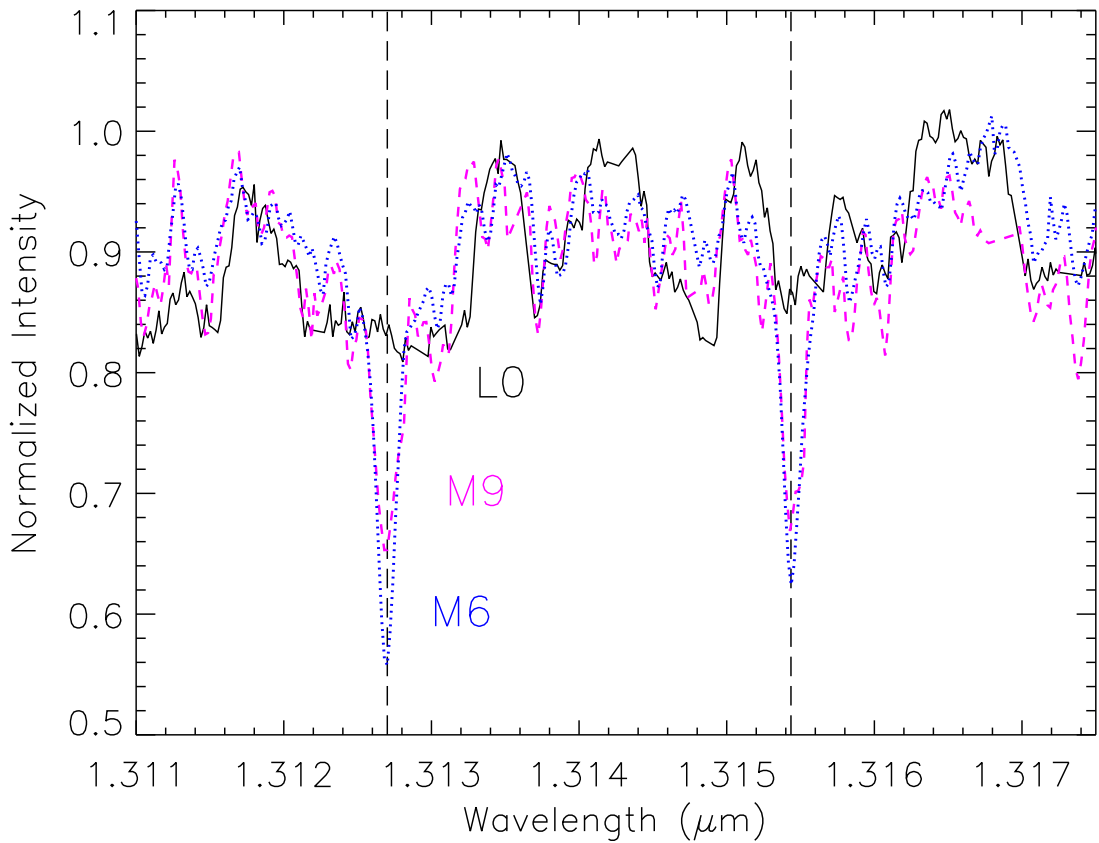} \caption{Part of the spectral
region covered by order 58 containing the Al I doublet at 1.3127
and 1.3154 \um is compared for the M6 (Wolf 359; dotted), M9 (2MASS
0140+27; dashed) and L0 (2MASS 0345+25; solid) objects shown in
Figures 1-3. The broad absorption features
are due to blended \h2o transitions. All spectra are normalized to
the same level and the intensity axis is plotted from 0.5 instead
of zero for clarity of presentation. The Al I lines remain
significant until M9 and then vanish abruptly between M9 to L0.}
\end{figure}
\epsscale{1}

\clearpage

\begin{figure}[!htp]
\includegraphics[scale=0.65,angle=90]{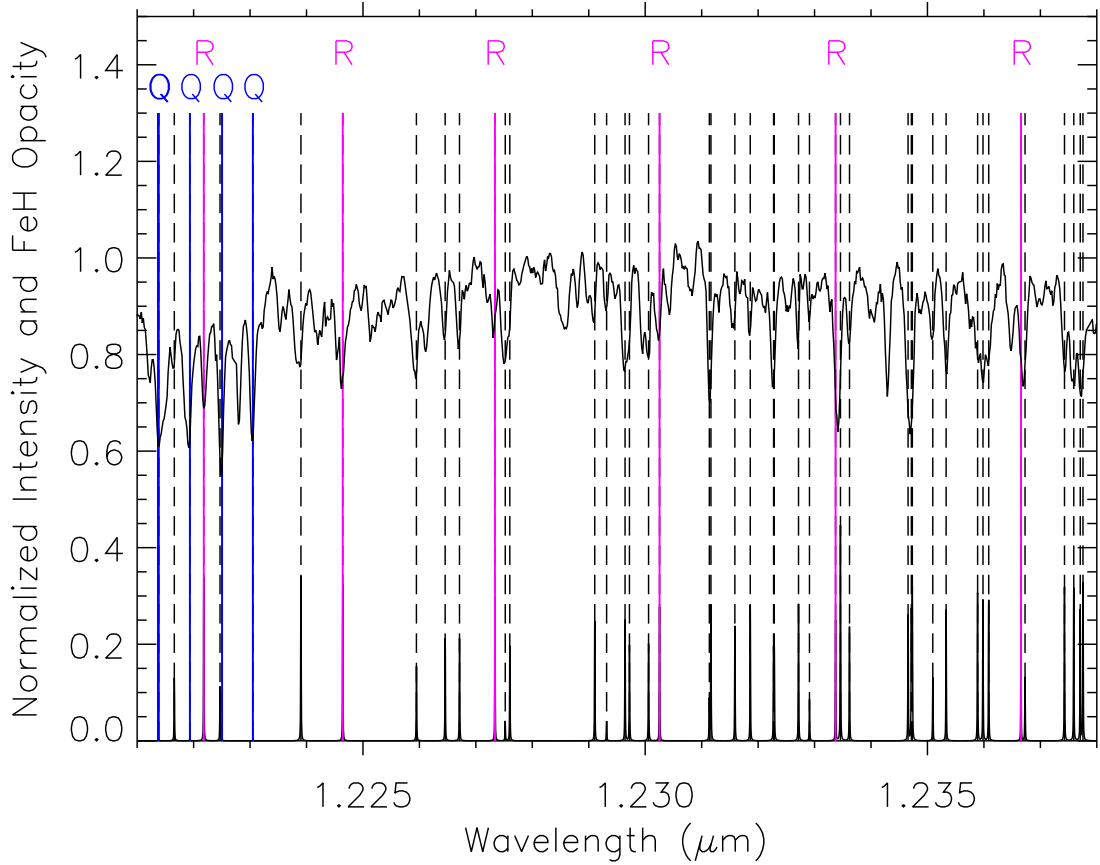} \caption{Order 62 from
1.221-1.238 $\mu$m for the M9 dwarf 2MASS 0140+27 is
over-plotted on a scaled FeH opacity plot smoothed to
R$\sim$20,000. The region near 1.222 $\mu$m previously attributed
to the Q-branch of the F$^{4}$ $\Delta_{7/2}$ - X$^{4}$
$\Delta_{7/2}$ system of FeH is resolved, and almost every other
feature is identified with P-branch (dotted vertical lines not
otherwise labeled) or R-branch (labeled) lines of the 0--1 band.
Lines not attributed to FeH are discussed in the text.}
\end{figure}
\epsscale{1}

\clearpage
\begin{figure}[!htp]
\includegraphics[scale=0.65,angle=90]{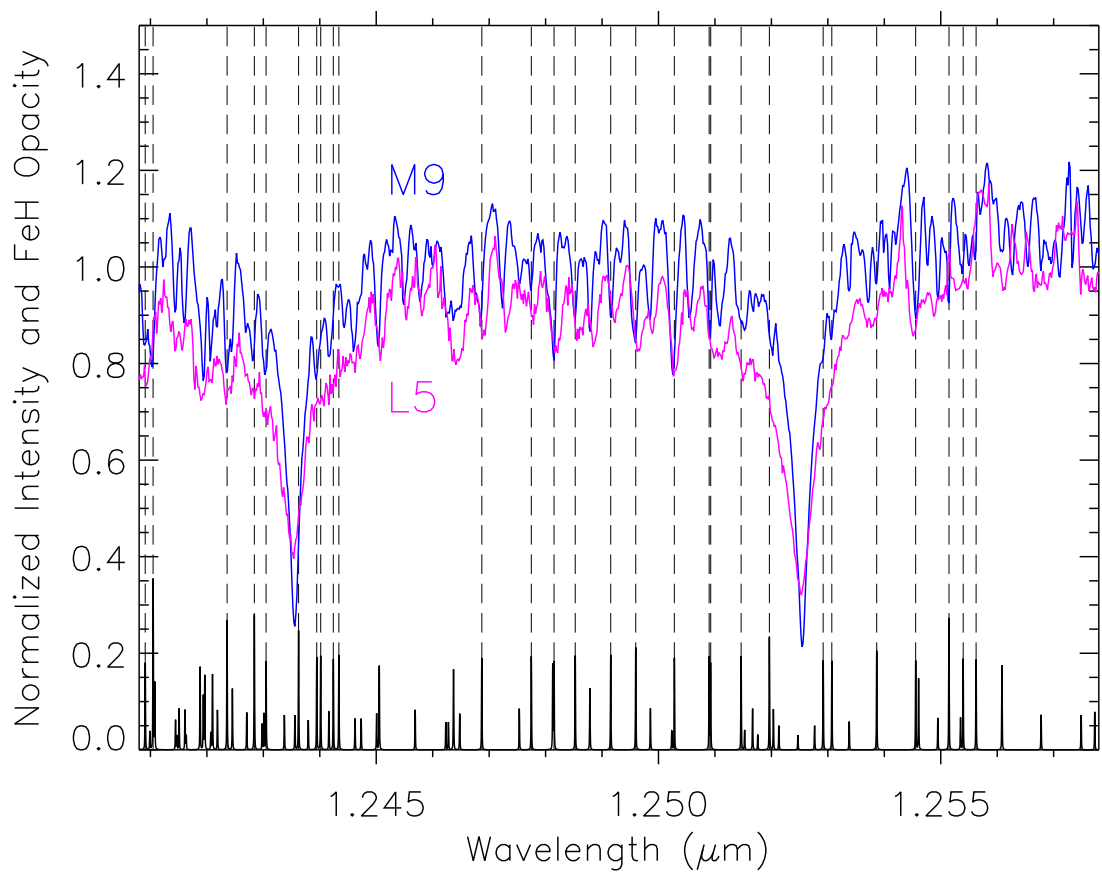} \caption{A plot of FeH opacity at
2000 K and a pressure of 1 bar, overlaid with the order 61
spectra of the M9 (2MASS 0140+27)  and L5 (2MASS 1507$-$16) dwarfs
from Table 1. Dashed lines show
the location of the FeH transitions and the darker solid lines
correspond to a resolution of R$\sim$20,000. Although the features
are broader in the L5 dwarf, almost all of the weaker features in
order 61 can be attributed to FeH. Lines not attributed to FeH are
discussed in the text. Note the broadening of the strong K I lines
from M9 to L5.}
\end{figure}
\epsscale{1}

\clearpage
\begin{figure}[!htp]
\includegraphics[scale=0.65,angle=90]{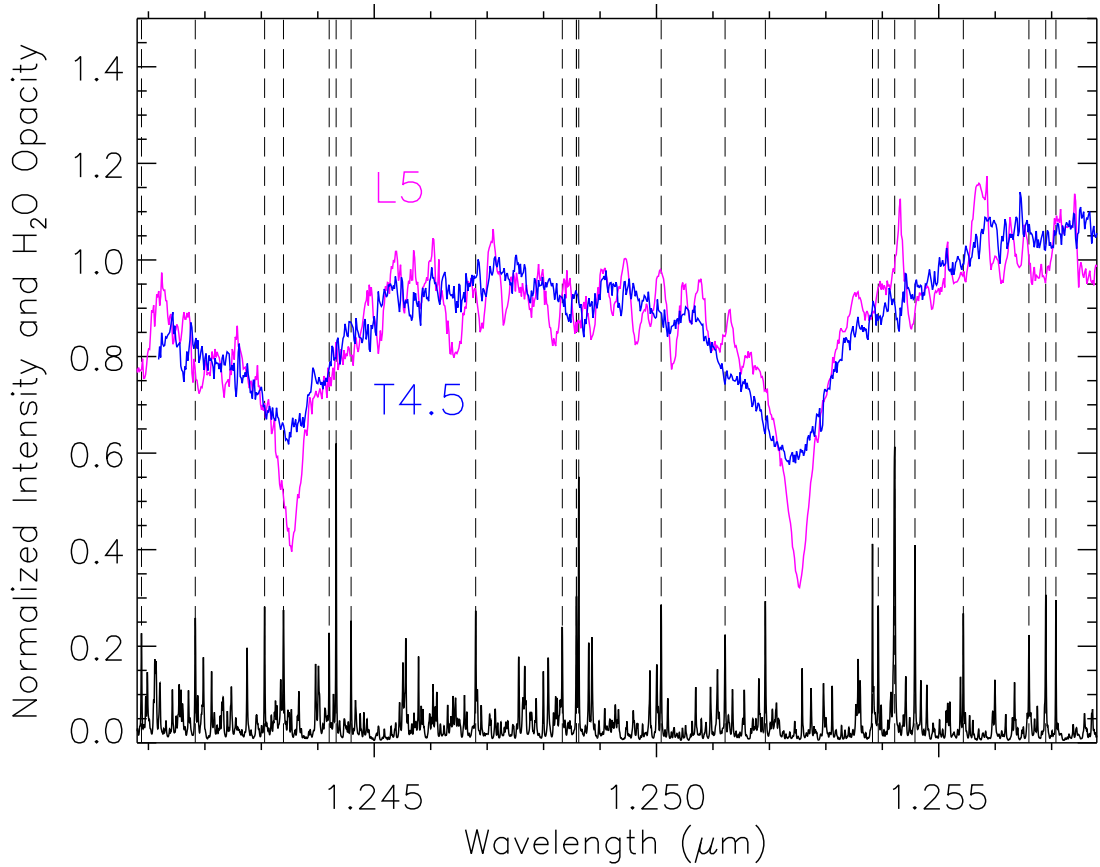} \caption{A plot of H$_{2}$O
opacity at 1000 K and a pressure of 1 bar, is overlaid with the
order 61 spectra of the L5 (2MASS 1507$-$16) and T4.5
(2MASS 0559$-$14) dwarfs from Table
1. Dashed lines show the location of the \h2o transitions and the
solid lines correspond to a resolution of R$\sim$20,000. H$_{2}$O
features do not dominate either spectrum. Spectral features in the
L5 that were not consistent with FeH cannot be attributed to
H$_{2}$O. FeH is almost completely absent from the T4.5.}
\end{figure}
\epsscale{1}

\clearpage
\begin{figure}[!htp]
\includegraphics[scale=0.65,angle=90]{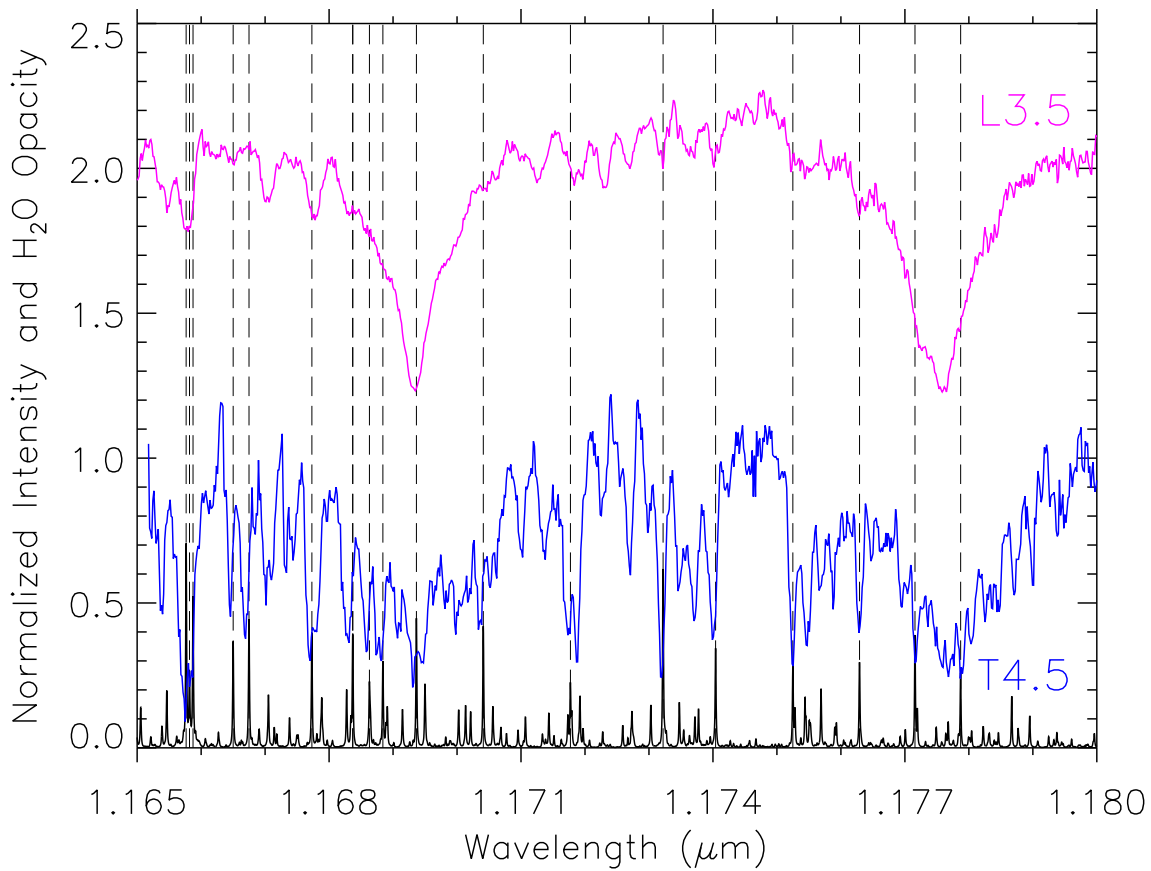} \caption{A plot of H$_{2}$O
opacity at 1000 K and a pressure of 1 bar, overlaid with the
order 65 NIRSPEC high-resolution spectra of the L3.5 (2MASS 0036+18)
and T4.5 (2MASS 0559$-$14) dwarfs
from Table 1. As in the previous figure, the solid lines
correspond to a resolution of R$\sim$20,000. No FeH lines are
listed in the opacity tables for this order. All of the strong,
sharper features in the T4.5 spectrum correlate well with H$_{2}$O
transitions. For the L3.5 the correlation is weaker but still
present. Note the broadening of the K I lines from L3.5 to T4.5. }
\end{figure}
\epsscale{1}

\clearpage
\thispagestyle{empty}
\begin{figure}[!htp]
\vspace*{-25mm}
\epsscale{0.95} \plotone{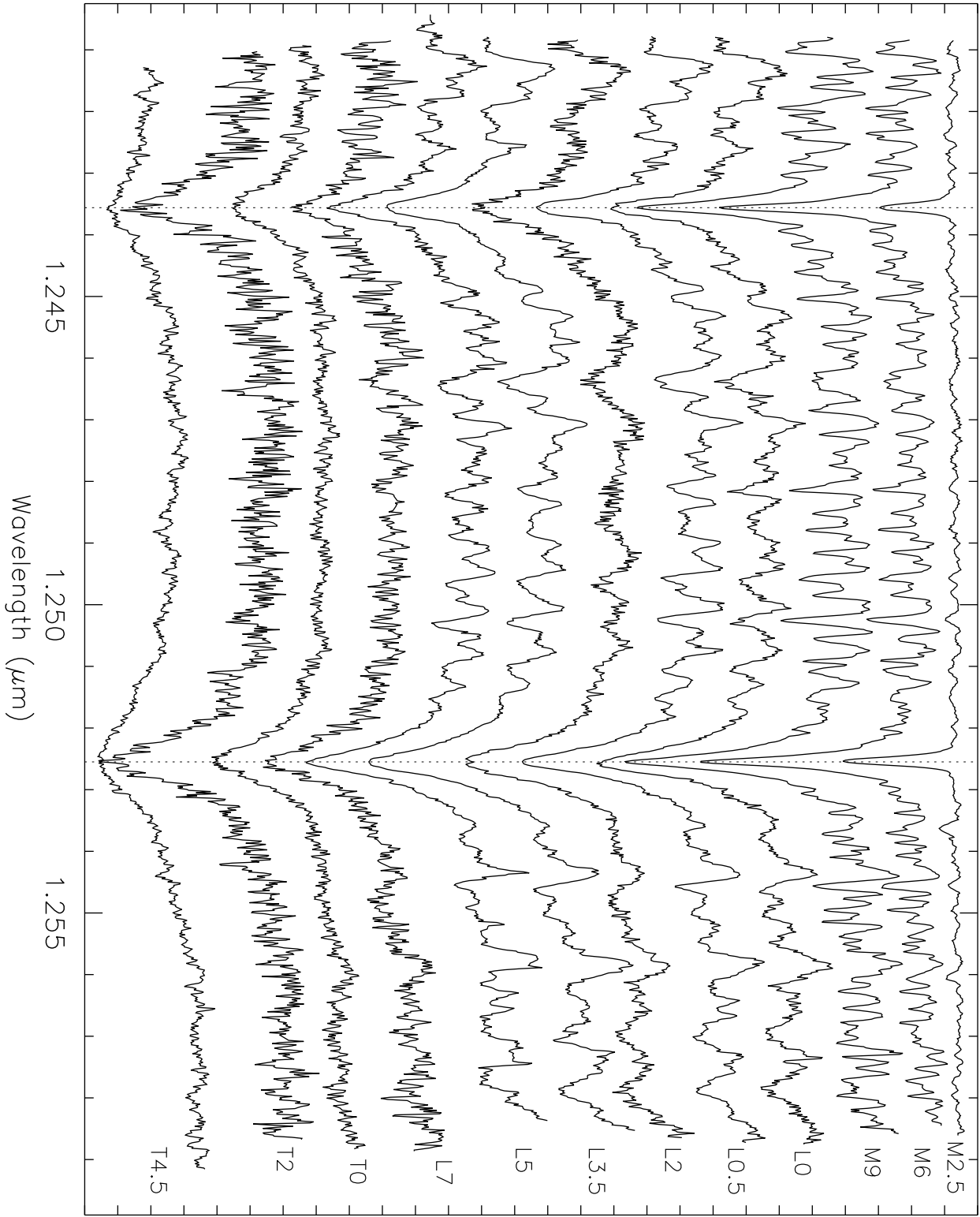} \caption{A spectral sequence
of NIRSPEC echelle order 61 (1.241-1.258 $\mu$m) containing the
longer wavelength K I doublet. Spectra are normalized to unity at
peak flux and offset by a constant. Each spectrum has been shifted
to the laboratory (vacuum) rest frame. The sequence of spectral
types corresponds to Table 1; from top to bottom the range is M2.5
to T4.5. Note the presence in the M2.5 object of a very slight
asymmetry in the position of the 1.252554 $\mu$m K I line that is
not seen in the other K I line. This is due to blending with a Cr
I line at 1.252527 $\mu$m, only 0.27\AA~ to the blue. }
\end{figure}
\epsscale{1}

\clearpage

\begin{figure}[!htp]
\epsscale{0.95} \plotone{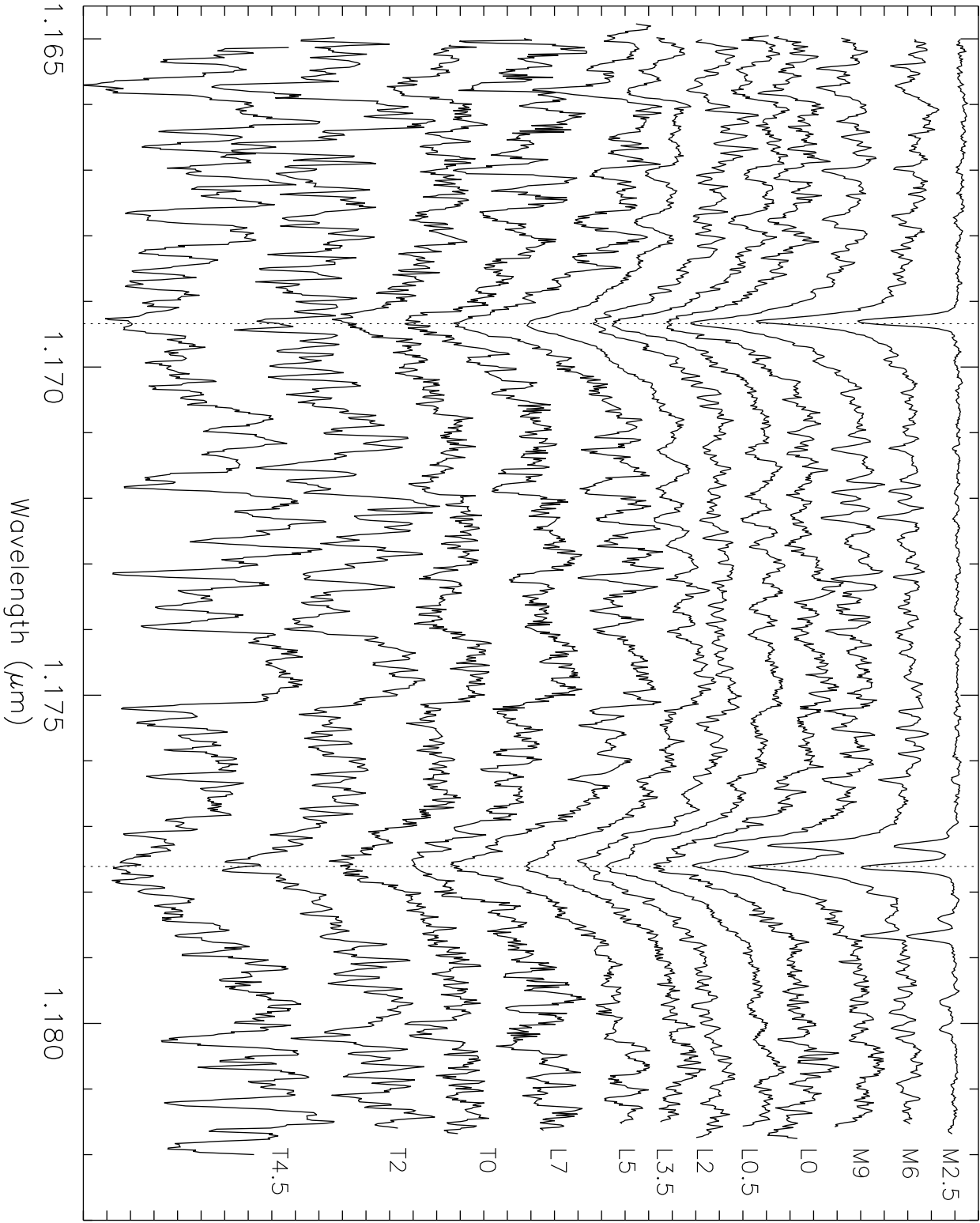} \caption{A spectral sequence
of NIRSPEC echelle order 65 (1.165-1.181 $\mu$m) containing the
shorter wavelength K I doublet. Spectra are normalized to unity
and offset by a constant. Each spectrum has been shifted to the
laboratory (vacuum) rest frame. The sequence of spectral types
corresponds to Table 1; from top to bottom the range is M2.5 to
T4.5.}
\end{figure}
\epsscale{1}

\clearpage
\begin{figure}[!htp]
\includegraphics[scale=0.8,angle=90]{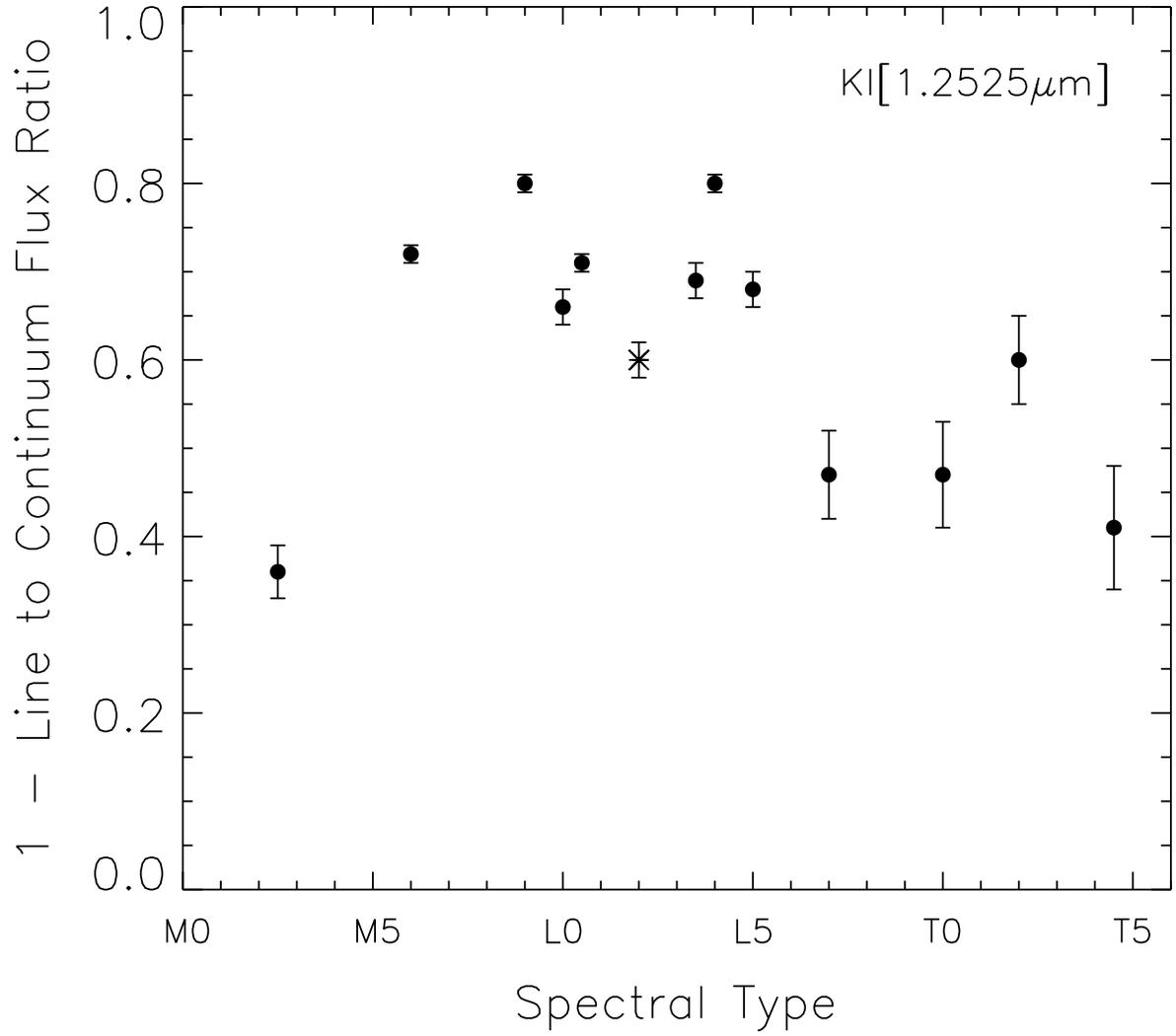}
  \caption{A plot of one minus the
line to continuum flux ratio is used to measure the depth of the
1.2525 $\mu$m K I line as a function of spectral type. For a very
strong line the index approaches unity. The object indicated by
the star symbol is Kelu-1AB.}
\end{figure}
\epsscale{1}

\clearpage
\begin{figure}[!htp]
\includegraphics[scale=0.8,angle=90]{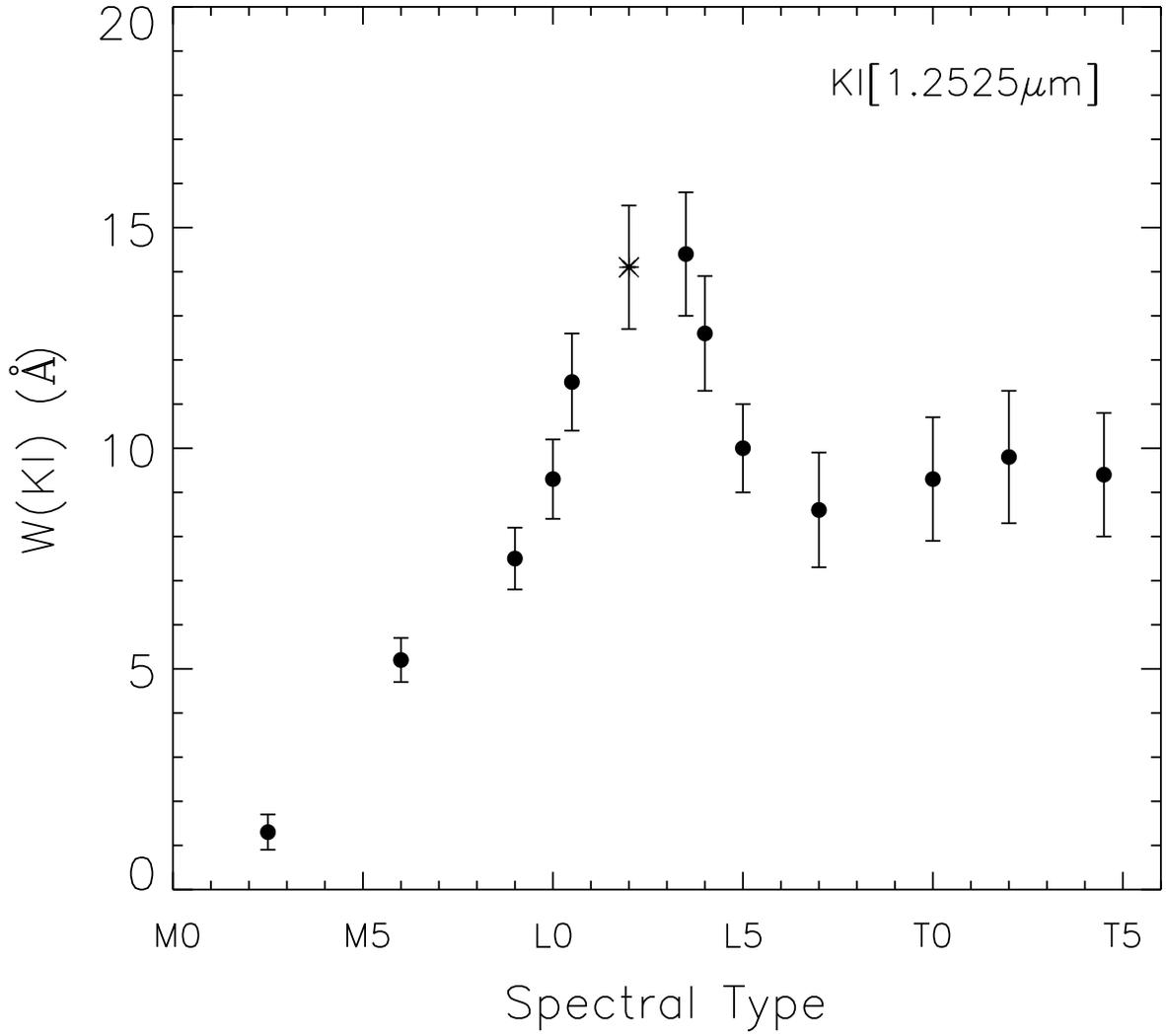} \caption{A plot of the equivalent
width (W, in \AA) of the 1.2525 $\mu$m K I line as a function of
spectral type. Error bars are estimated from a sequence of trial
fits to the continuum. The object indicated by the star symbol is
Kelu-1AB.}
\end{figure}
\epsscale{1}

\clearpage
\begin{figure}[!htp]
\includegraphics[scale=0.8,angle=90]{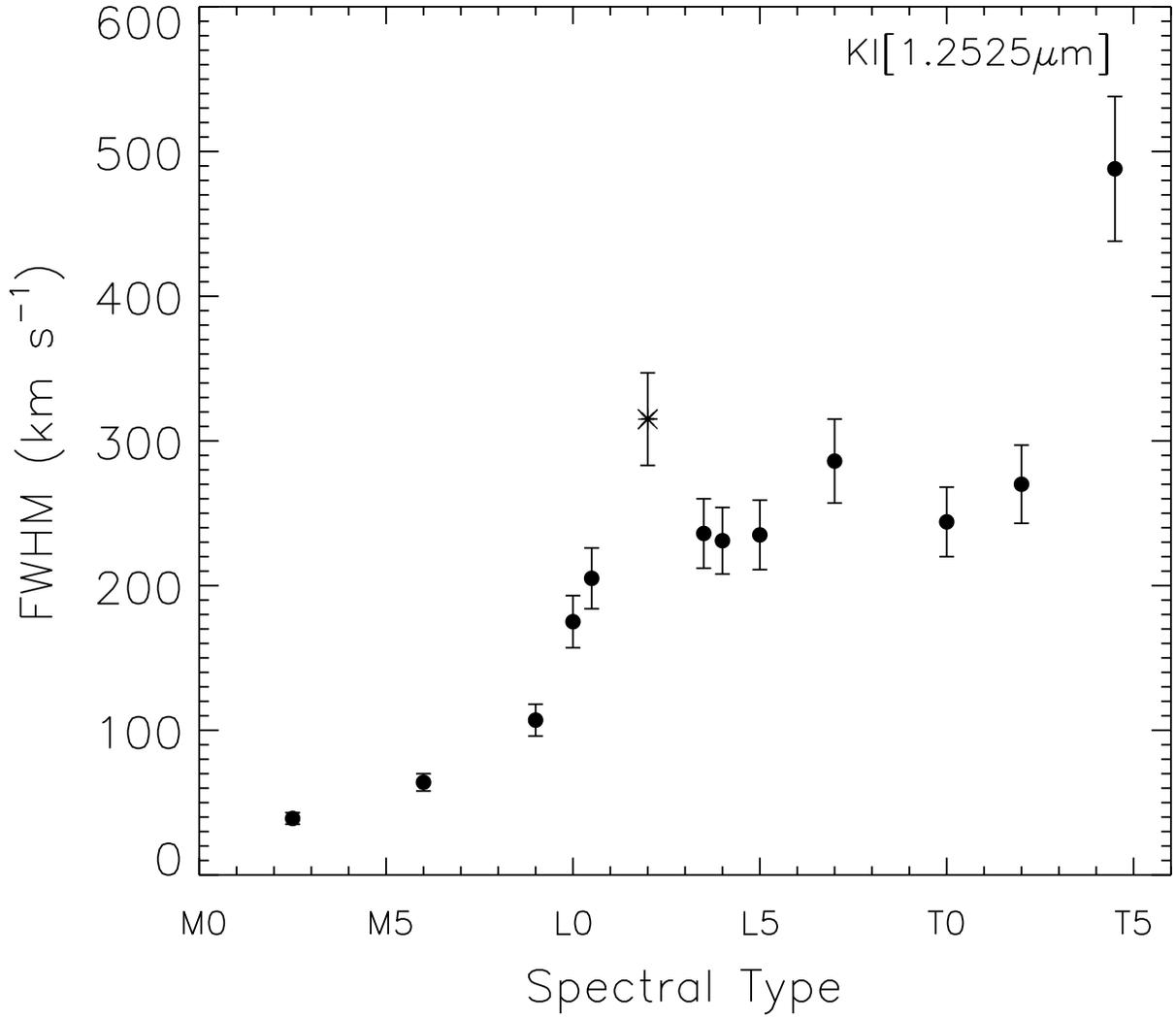} \caption{A plot of the FWHM (in
km s$^{-1}$) of the 1.2525 $\mu$m K I line as a function of
spectral type. The object indicated by the star symbol is Kelu-1AB.}
\end{figure}
\epsscale{1}

\clearpage
\begin{figure}[!htp]
\includegraphics[scale=0.8,angle=90]{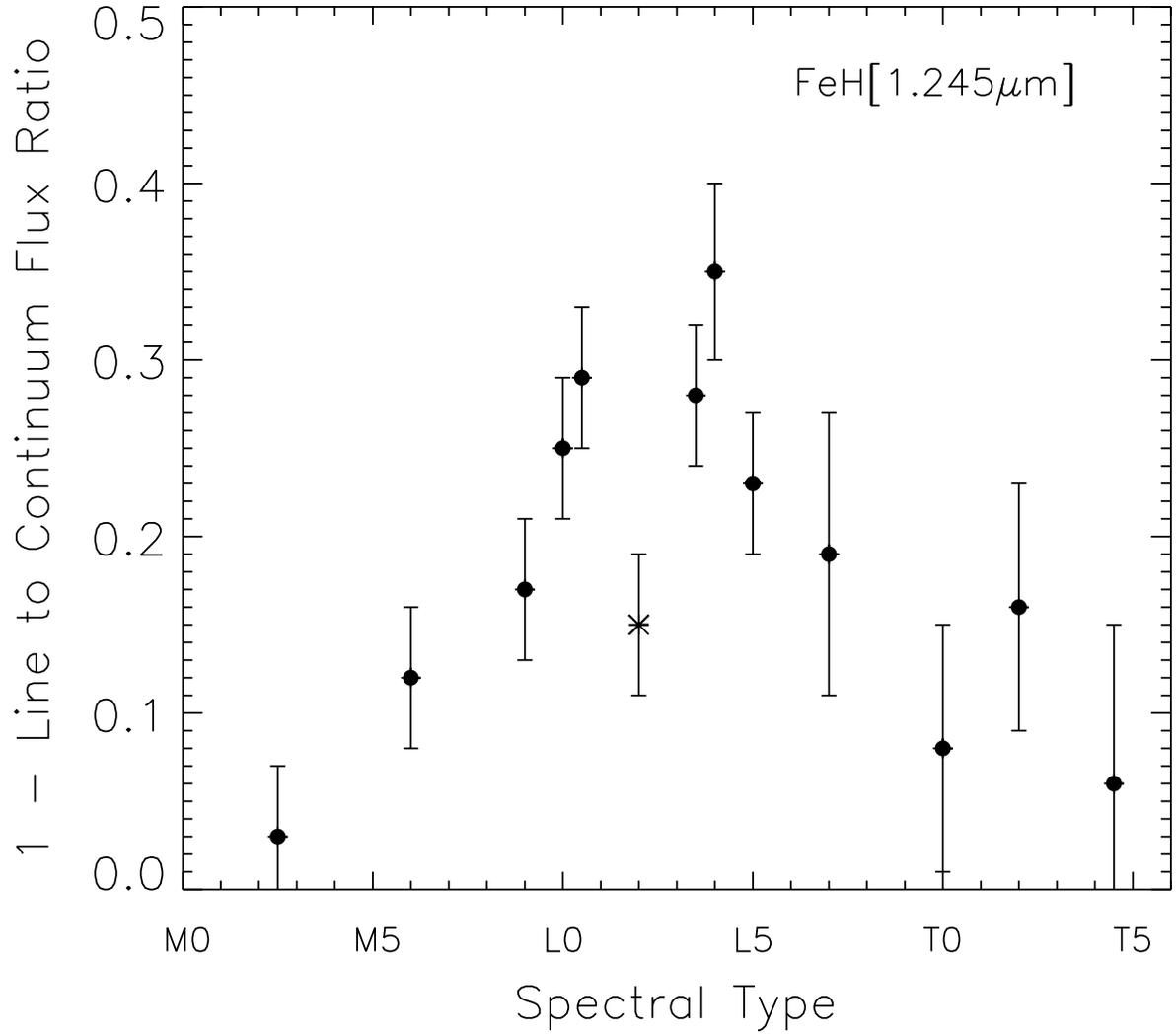} \caption{A plot of one minus the
line to continuum flux for the strong FeH feature at 1.245 $\mu$m
as a function of spectral type. Again, the object indicated by the
star symbol is Kelu-1AB which appears to be anomalous.}
\end{figure}
\epsscale{1}

\clearpage
\begin{figure}[!htp]
\includegraphics[scale=0.8,angle=90]{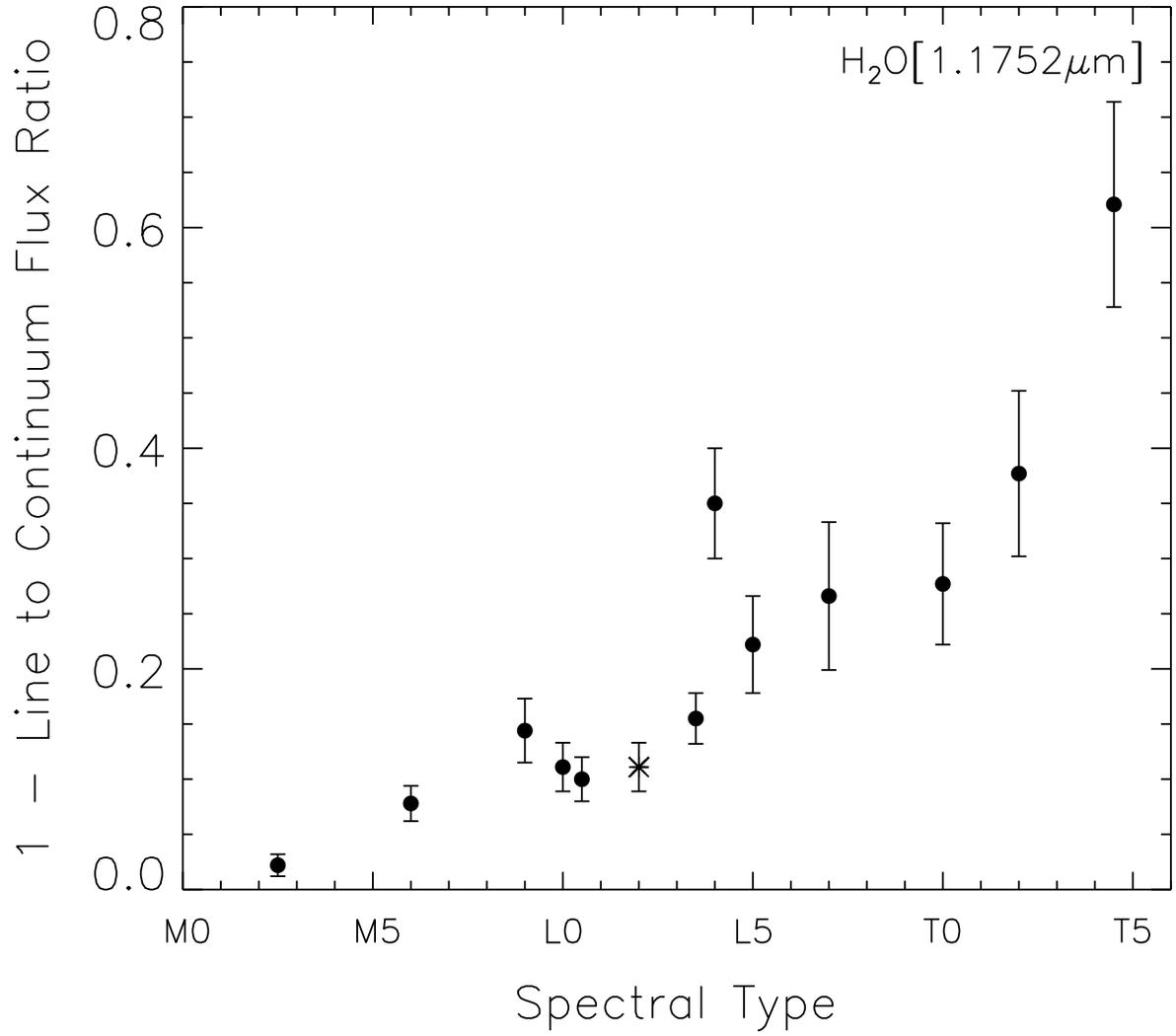} \caption{A plot of one minus the
line to continuum flux ratio for the strong \h2o feature at 1.1752
$\mu$m as a function of spectral type. As before, the object
indicated by the star symbol is Kelu-1AB. In this plot it is GD165B
(L4) that appears slightly above the general trend.}
\end{figure}
\epsscale{1}

\clearpage

\begin{figure}[!htp]
\includegraphics[scale=1.0,angle=0]{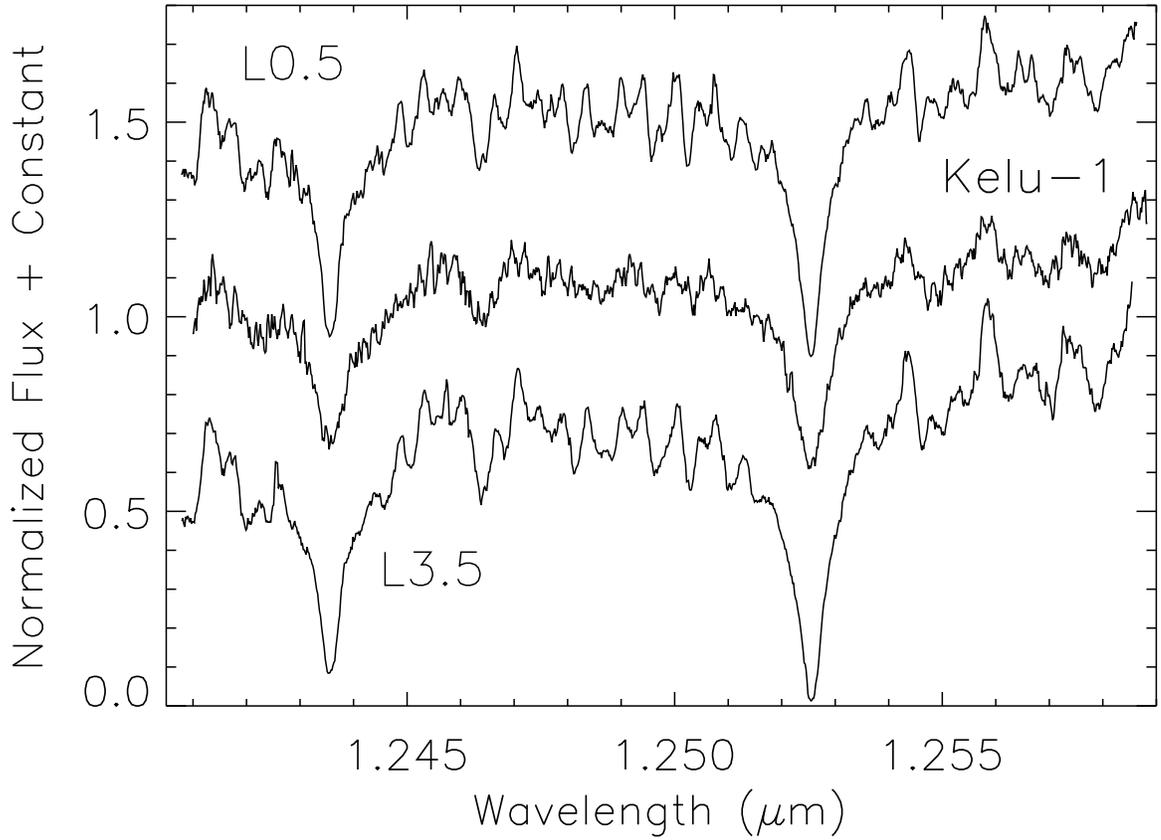} \caption{The high resolution
spectrum at 1.25 $\mu$m (order 61) of the L2 dwarf Kelu-1AB compared
to L0.5 and L3.5 dwarfs (2MASS 0746+20AB and 2MASS 0036+18,
respectively). The K I lines are significantly broader
and shallower in Kelu-1AB, even compared to the binary L0.5. Spectra are normalized to unity and
offset by a constant. Each spectrum has been shifted to the
laboratory (vacuum) rest frame. }
\end{figure}
\epsscale{1}

\clearpage

\begin{figure}[!htp]
\epsscale{0.95} \plotone{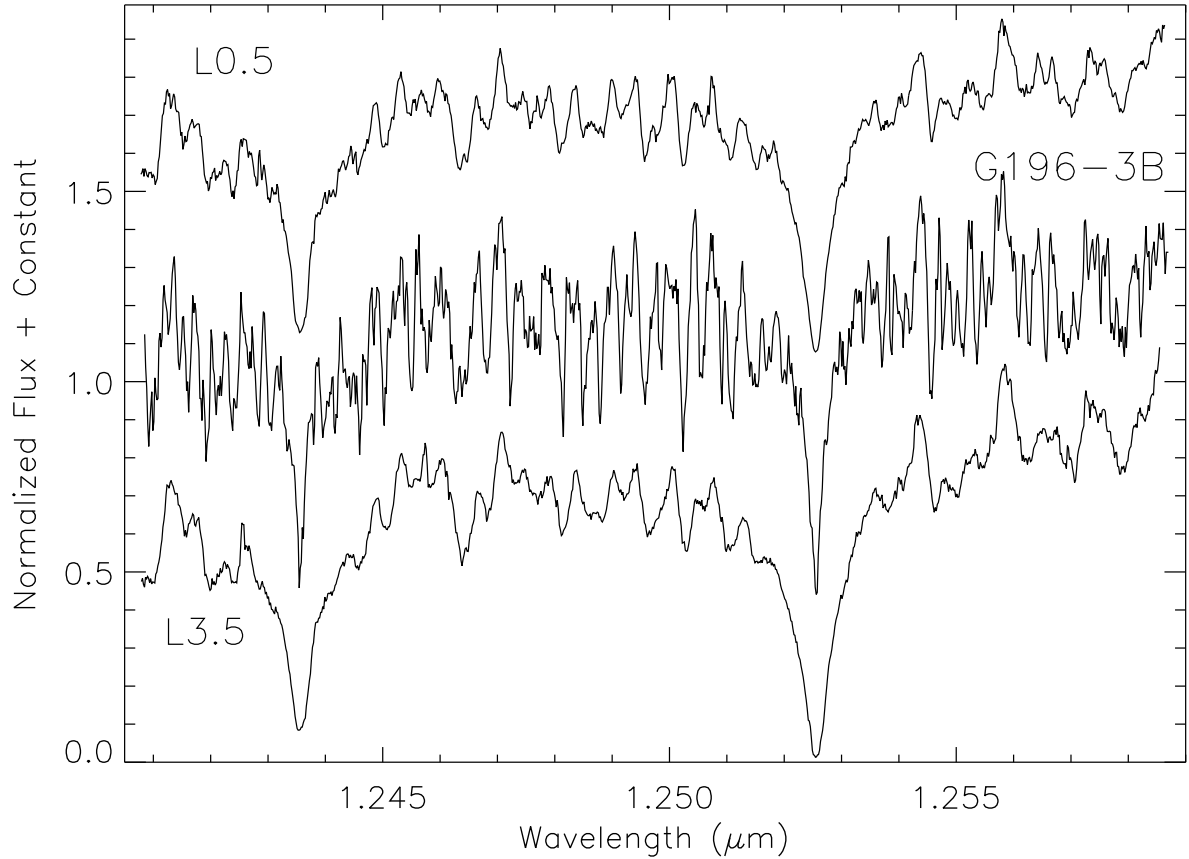} \caption{The high resolution
spectrum at 1.25 $\mu$m (order 61) of the young L2 companion
G196-3B compared to field L0.5 and L3.5 dwarfs (2MASS 0746+20AB and 2MASS 0036+18,
respectively). The K I lines are
significantly weaker and narrower. Spectra are normalized to unity
at peak flux and offset by a constant. Each spectrum has been
shifted to the laboratory (vacuum) rest frame. }
\end{figure}
\epsscale{1}

\clearpage
\begin{figure}[!htp]
\epsscale{0.95} \plotone{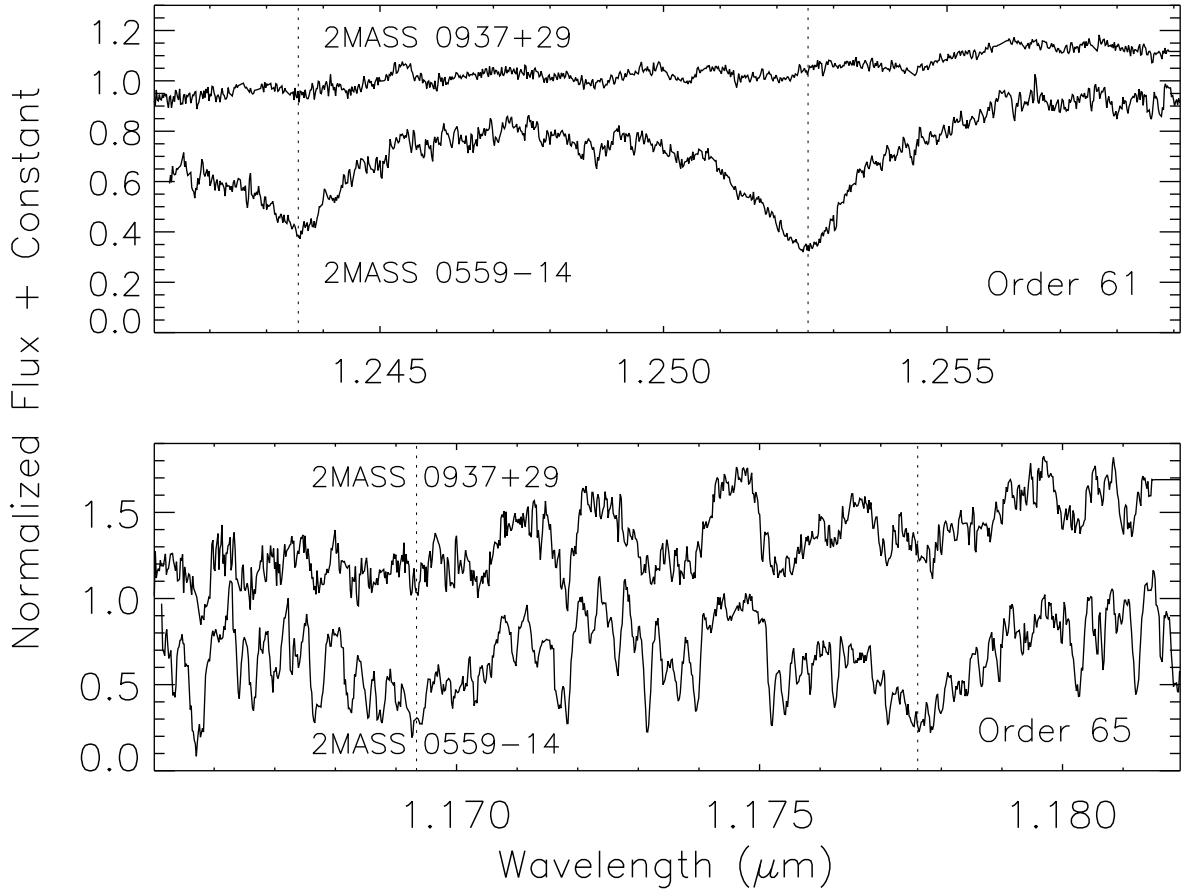} \caption{The high resolution
spectra at 1.25 $\mu$m (order 61) and 1.173 $\mu$m (order 65) of
the peculiar T6 dwarf 2MASS 0937+29 compared to that of the T4.5
dwarf 2MASS 0559$-$14. The K I lines are completely absent in both
orders. Transitions due to $H_{2}$O in order 65 are present but
muted in the peculiar T6.}
\end{figure}
\epsscale{1}

\end{document}